\documentclass[fleqn,usenatbib]{mnras}

\usepackage{newtxtext,newtxmath}

\usepackage[T1]{fontenc}

\DeclareRobustCommand{\VAN}[3]{#2}
\let\VANthebibliography\thebibliography
\def\thebibliography{\DeclareRobustCommand{\VAN}[3]{##3}\VANthebibliography}

\usepackage{graphicx}	
\usepackage{amsmath}	
\usepackage{adjustbox}
\usepackage{subcaption}
\usepackage{color}

\title[Cosmological tensions from high-$z$ cosmography]{Tensions with the flat $\boldsymbol{\Lambda}$CDM model from high-redshift cosmography}

\author[G. Bargiacchi et al.]{
G. Bargiacchi,$^{1,2}$\thanks{Corresponding author, E-mail: giada.bargiacchi@unina.it}
M. G. Dainotti,$^{3,4,5}$
S. Capozziello$^{1,2,6}$
\\
$^{1}$Scuola Superiore Meridionale, Largo S. Marcellino 10, 80138 Napoli, Italy \\
$^{2}$Istituto Nazionale di Fisica Nucleare (INFN), Sez. di Napoli, Complesso Univ. Monte S. Angelo, Via Cinthia 9, 80126, Napoli, Italy\\
$^{3}$ National Astronomical Observatory of Japan, 2 Chome-21-1 Osawa, Mitaka, Tokyo 181-8588, Japan\\
$^{4}$ The Graduate University for Advanced Studies, SOKENDAI, Shonankokusaimura, Hayama, Miura District, Kanagawa 240-0193, Japan\\
$^{5}$ Space Science Institute, 4765 Walnut St, Suite B, 80301 Boulder, CO, USA\\
$^{6}$Dipartimento di Fisica "E. Pancini", Università degli Studi di Napoli Federico II, Complesso Univ. Monte S. Angelo, Via Cinthia 9, 80126, Napoli, Italy
}

\date{Accepted XXX. Received YYY; in original form ZZZ}

\pubyear{2023}

\begin{document}
\label{firstpage}
\pagerange{\pageref{firstpage}--\pageref{lastpage}}
\maketitle

\begin{abstract}
The longstanding search for the cosmological model that best describes the Universe has been made more intriguing since the recent discovery of the Hubble constant, $H_{0}$, tension observed between the value of $H_{0}$ from the Cosmic Microwave Background and from type Ia supernovae (SNe Ia). Hence, the commonly trusted flat $\Lambda$CDM model is under investigation.
In this scenario, cosmography is a very powerful technique to investigate the evolution of the Universe without any cosmological assumption, thus revealing tensions between observational data and predictions from cosmological models in a completely model-independent way.
We here employ a robust cosmographic technique based on an orthogonal logarithmic polynomial expansion of the luminosity distance to fit quasars (QSOs) alone and  QSOs combined with Gamma-Ray Bursts (GRBs), SNe Ia, and Baryon Acoustic Oscillations. To apply QSOs and GRBs as probes we use, respectively, the Risaliti-Lusso relation between ultraviolet and X-ray luminosities and the ``Dainotti GRB 3D relation"
among the rest-frame end time of the X-ray plateau emission, its corresponding luminosity, and the peak prompt luminosity. We also correct QSOs and GRBs for selection biases and redshift evolution and we employ both the traditional Gaussian likelihood and the newly discovered best-fit likelihoods for each probe investigated. This comprehensive analysis reveals a strong tension ($> 4 \, \sigma$) between our data sets and the flat $\Lambda$CDM model proving the power of both the cosmographic approach and high-redshift sources, such as QSOs and GRBs, which can probe the Universe at early epochs.
\end{abstract}




\section{Introduction}
\label{intro}

Cosmological probes such as type Ia supernovae (SNe Ia; \citealt{riess1998,perlmutter1999}) and the combination of SNe Ia, Baryonic Acoustic Oscillations (BAOs), Cosmic Microwave Background (CMB), and lensing \citep{planck2018} have proved that the Universe is currently in a phase of accelerated expansion, which is commonly ascribed to a dark energy component. According to the $\Lambda$ cold dark matter (CDM) model ($\mathrm{\Lambda}$CDM), this dark energy component is a cosmological constant ($\mathrm{\Lambda}$) originated by vacuum quantum fluctuations \citep{2000IJMPD...9..373S,2001LRR.....4....1C,2003RvMP...75..559P}. Nevertheless, the discrepancy between the current theoretical and observed value of $\Lambda$ poses the cosmological constant problem (\citealt{1989RvMP...61....1W}). This problem historically questions the $\Lambda$CDM model, together with other well-known theoretical issues, such as the fine-tuning and the coincidence problems.
On the other hand, the flat $\mathrm{\Lambda}$CDM model can well reproduce several observations from SNe Ia, BAOs, and CMB. This interesting puzzle regarding the flat $\Lambda$CDM model has been recently made more intriguing by the discovery of the Hubble constant, $H_{0}$, tension. Indeed, the $H_0$ value inferred from the CMB, assuming a flat $\mathrm{\Lambda}$CDM model \citep[$H_0 = 67.4 \pm 0.5  \, \mathrm{km} \, \mathrm{s}^{-1} \, \mathrm{Mpc}^{-1}$,][]{planck2018}, is incompatible, at $\geq 4 \, \sigma$ with the value measured locally from Cepheids and SNe Ia \citep[$H_0 = 73.04 \pm 1.04  \, \mathrm{km} \, \mathrm{s}^{-1} \, \mathrm{Mpc}^{-1}$,][]{ 2022ApJ...934L...7R}, and with that reported in \citet{2020MNRAS.498.1420W} obtained from lensed quasars ($H_0 = 73.3 \pm 1.7  \, \mathrm{km} \, \mathrm{s}^{-1} \, \mathrm{Mpc}^{-1}$). 
Other cosmological probes have been further employed to shed light on this tension showing even a more complex scenario. For example, cosmic chronometers tend to a $H_0$ value close to the one of the CMB \citep{2018JCAP...04..051G}, while time-delay and strong lensing of quasars (QSOs) prefer the one of SNe Ia \citep{2019ApJ...886L..23L}, and other probes, such as QSOs \citep{biasfreeQSO2022}, the Tip of the Red-Giant Branch (TRGB; \citealt{2021ApJ...919...16F}), and Gamma-Ray Bursts (GRBs; \citealt{cardone09,Dainotti2013a, Postnikov14,Dainotti2021ApJ...912..150D,2022Galax..10...24D,Dainotti2022MNRAS.tmp.2639D,Dainotti2022PASJ...74.1095D}) show a $H_0$ value halfway between the one of SNe Ia and the one of CMB.
Many theoretical works have been discussed to explain this tension \citep{DiValentino:2020hov,Dainotti2021ApJ...912..150D,Schiavone:2022wvq}.

In this framework, techniques which are completely independent from the cosmology are among the most powerful tools that have been so far proposed to unveil the actual structure of the Universe without any cosmological assumption. In this regard, cosmography \citep{1972gcpa.book.....W} is one of the most consolidated approaches.
It relies only on the isotropy and homogeneity of the Universe by assuming the Friedmann-Lema\^{i}tre-Robertson-Walker (FLRW) metric. As a consequence, cosmography provides a description of the Universe which is purely geometrical and where all the physics is embedded in the scale factor $a(t)$ of the FLRW metric. Nevertheless, it does not need an explicit expression for $a(t)$, but only an analytical form for it. Hence, this technique supplies a completely model-independent way to investigate the evolution of the Universe.
The main advantage of this approach is that, if the investigated cosmographic parameterization has a number of free parameters that makes it sufficiently flexible, it can fit the data with high accuracy, and then the obtained cosmographic parameters can be used to test any cosmological model \citep{2018MNRAS.476.3924C,2019IJMPD..2850154E}. More precisely, if we expand the considered cosmological model and the cosmographic expansion, we obtain the relations between the free cosmological and cosmographic parameters \citep{2014PhRvD..90d3531A,2019IJMPD..2830016C,Benetti}, see also Section \ref{comparison}. These relations allows us to compare the best-fit of the cosmographic model with the prediction of the investigated cosmological model \citep[see e.g.][]{2004ApJ...607..665R}.
Nonetheless, the cosmographic approach presents one major issue: as in cosmological studies, high-redshift (i.e. $z>2$) sources are essential to investigate possible deviations from the theoretical predictions of cosmological models, but in the high-redshift interval convergence issues may emerge in the cosmographic expansion \citep[see e.g.][]{2020MNRAS.494.2576C}. The traditional Taylor expansion commonly used in cosmography to analytically approximate $a(t)$ \citep{2004CQGra..21.2603V,2019IJMPD..2830016C} is an example of this problem. Indeed, following the definition of the Taylor expansion, it gives an approximate expression of the luminosity distance $D_{L}(z)$ around $z=0$, and thus can be applied only in the limit $z\leq1$ and not to high-redshift probes, such as QSOs and GRBs, observed up to $z = 7.64$ \citep{2021ApJ...907L...1W} and $z=9.4$ \citep{2011ApJ...736....7C}, respectively. 
Moreover, also SNe Ia have been recently observed up to $z=2.26$ \citep{Rodney} by the \textit{Pantheon} survey \citep{scolnic2018}, which marks the problem of applying the Taylor expansion in cosmography. 

While SNe Ia are commonly trusted as the most reliable and powerful standard candles in cosmology, QSOs and GRBs have only been recently promoted to cosmological tools, attracting more and more the interest of the cosmological community due to their invaluable role at high redshifts.
To employ QSOs as cosmological probes, we use the X-ray to ultraviolet (UV) relation (X-UV relation), also called Risali-Lusso relation (RL), between the UV and X-ray luminosities. This correlation was proposed after the first X-ray surveys
\citep{1979ApJ...234L...9T,1981ApJ...245..357Z,1986ApJ...305...83A}, and then validated with different QSO samples \citep[e.g.][]{steffen06,just07,2010A&A...512A..34L,lr16,2021A&A...655A.109B}. This relation is physically supported by the commonly accepted theoretical model in which the emission of QSOs is supplied by the accretion on a central supermassive black hole, which converts mass into energy with high efficiency \citep{1998A&A...334...39S,1999RvMPS..71..180H,qsophysics,2020MNRAS.498.5652K}. The RL relation has been successfully employed for cosmological applications \citep{rl15,rl19,2020A&A...642A.150L,2021A&A...649A..65B,2022MNRAS.515.1795B,DainottiQSO,2020MNRAS.492.4456K,2020MNRAS.497..263K,2021MNRAS.502.6140K,Colgain2022arXiv220611447C,2022MNRAS.510.2753K,2022arXiv220310558C,2022arXiv221014432W,2022MNRAS.517.1901L,
biasfreeQSO2022,Bargiacchi2023arXiv230307076B,Dainotti2023arXiv230510030D,DainottiGoldQSOApJ2023}. The reliability of this relation for cosmological purposes has also been recently proved by \citet{DainottiQSO}, which have accounted for selection biases and redshift evolution of QSO luminosities.  
Analogously, GRBs can be standardized as cosmological tools through the GRB fundamental plane relation, also named the three-dimensional (3D) Dainotti relation, which is an implementation of the previous two-dimensional relation, among the rest-frame time at the end of the plateau emission, $T^{*}_{\chi}$, the corresponding X-ray luminosity of the light curve, $L_{\chi}$ \citep{Dainotti2008,Dainotti2010ApJ...722L.215D,dainotti11a,Dainotti2017A&A...600A..98D}, and the prompt peak luminosity, $L_{peak}$ \citep{Dainotti2016ApJ...825L..20D,Dainotti2017ApJ...848...88D,Dainotti2020a,Srinivasaragavan2020}. 
This relation is theoretically supported by the magnetar model \citep{2001ApJ...552L..35Z,2011A&A...526A.121D, 2012MNRAS.425.1199B,Rowlinson2014MNRAS.443.1779R,Rea2015ApJ...813...92R}.
The two-dimensional Dainotti relation and the fundamental plane relation have already been applied in several cosmological studies \citep{cardone09,cardone10,Postnikov14,Dainotti2013a,Dainotti2022PASJ,Dainotti2022MNRAS.514.1828D,Dainotti2022MNRAS.tmp.2639D,Bargiacchi2023arXiv230307076B, Dainotti2023arXiv230510030D}.
The two-dimensional Dainotti relation has also been discovered in optical \citep{Dainotti2020b, Dainotticlosureoptical2022ApJ...940..169D} and radio
\citep{Levine2022ApJ...925...15L}, while the 3D correlation has also been found in high energy $\gamma$-rays \citep{Dainotti2021ApJS..255...13D} and in optical \citep{Dainotti2022ApJS..261...25D}. For a more extensive discussion on the GRB correlations
see also \citet{Dainotti2017NewAR..77...23D}, \citet{Dainotti2018PASP..130e1001D}, and  \citet{Dainotti2018AdAst2018E...1D}.

The combination of traditional low-redshift cosmological probes, such as SNe Ia and BAOs, and promising high-redshift sources, such as QSOs and GRBs, allows us
to effectively unveil tensions between the observations and the standard cosmological model through the cosmographic approach (see e.g. \citealt{2020ApJ...900...70R}). However,  this is possible only if the above-described convergence issues are solved. To this end, some techniques have been proposed, as for example rational polynomials such as the Pad\'e and
Chebyshev polynomials (see e.g. \citealt{2014PhRvD..90d3531A,2018MNRAS.476.3924C,2017A&A...598A.113D,2020JCAP...12..007Z}).
In this paper, we make use of a cosmographic technique based on the expansion of the luminosity distance in orthogonal polynomials of logarithmic functions. The robustness  and reliability of this approach has already been discussed in detail and proved in \citet{2021A&A...649A..65B}. This cosmographic model, indeed, completely overcomes the weakness of the Taylor expansion at high redshifts providing a suitable analytical reproduction of observational data and cosmological models to test physical models.
We stress that here the cosmographic approach is applied starting from QSOs as the major cosmological player.

We here leverage the power of this cosmographic technique to separately fit QSOs alone and QSOs joint with GRBs, SNe Ia from the \textit{Pantheon +} release, and BAOs to test the flat $\Lambda$CDM model. This allows us to unveil and quantify tensions between this cosmological model and the considered observational data through a completely model-independent technique. We also compare different treatments for the correction for redshift evolution of luminosities of QSOs and GRBs, and different likelihoods to show to what extent these choices affect the results.  
Compared to previous cosmographic analyses in which QSOs have been employed, this work manifestly presents several points of novelty. It is indeed the first time in the cosmographic realm that the most updated sample of QSOs is used in the whole redshift range of observations, including also low-redshift sources, as standalone probe and also combined with the latest samples of GRBs and SNe Ia. Moreover, the correction for selection biases and redshift evolution and the application of likelihoods different from the Gaussian one, which are crucial to properly and better estimate the parameters, have never been implemented in cosmography before.

The manuscript is structured as follows. Section \ref{data} describes the samples of QSOs, GRBs, SNe Ia, and BAO. Section \ref{methods} introduces the cosmographic model, how it can be compared to the predictions of the flat $\Lambda$CDM model, the physical quantities investigated for each probe, the technique applied to account for selection biases and correct for redshift evolution, and the fitting procedure with different likelihoods. In Section \ref{results}, we outline our results, also compared with previous works. Finally, in Section \ref{conclusions} we summarize our conclusions.

\section{Data sets}
\label{data}

In this work, we use both QSOs alone and QSOs combined with GRBs, SNe Ia from the \textit{Pantheon+} sample, and BAOs.
The QSO data set is the one composed of 2421 sources described in detail in \citet{2020A&A...642A.150L}. This sample has been built combining eight different sub-samples present in archives and in the
literature and is the most updated QSO catalog specifically released for cosmological purposes. Indeed, these sources have been carefully selected based on several criteria both in X-ray and UV bands (e.g. signal-to-noise ratio, the host galaxy contamination in UV, the absorption in X-ray, the Eddington bias) to remove as many as possible observational biases \citep[see][for details]{2020A&A...642A.150L}. 
Despite \citet{2020A&A...642A.150L}, as in other recent works \citep{DainottiQSO,biasfreeQSO2022,Bargiacchi2023arXiv230307076B,Dainotti2023arXiv230510030D,DainottiGoldQSOApJ2023}, we here employ all the 2421 sources which cover the redshift range $z=0.009 - 7.54$ to avoid imposing an arbitrary truncation in the redshift distribution of the sources and to leverage also the power of low-redshift QSOs to constrain parameters.

As GRB data set, we employ the ``Platinum" sample \citep{Dainotti2020ApJ...904...97D}, which has been further improved compared to the previous golden sample \citep{Dainotti2016ApJ...825L..20D,Dainotti2017ApJ...848...88D, Dainotti2020ApJ...904...97D}. It consists of 50 X-ray observations ranging from $z=0.055$ to $z=5$ which verify the following requirements: a) the plateau must have an inclination $<$41°, 
b) it must present at least 5 points in its beginning to allow a determination of the onset,
c) it must not show flares,
d) and it must lasts $> 500$s \citep[see also][]{Dainotti2022MNRAS.tmp.2639D,Dainotti2022MNRAS.514.1828D}.
This selection leads to the Platinum sample of 50 sources starting from 222 GRBs with a known redshift observed from January 2005 to August 2019 by the {\it Neil Gehrels Swift Observatory} (Swift). These data are taken from the Swift Burst Analyser \citep{Evans2009}, which includes (BAT), X-Ray Telescope (XRT), and optical (UVOT) repositories. These GRBs possess a plateau that can be fitted applying the \cite{2007ApJ...662.1093W} model.

For SNe Ia, we use the most up-to-date sample from the \textit{Pantheon+} release \citep{pantheon+}, which has improved the previous \textit{Pantheon} sample \citep{scolnic2018} with additional low-redshift observations. It is indeed composed of 1701 sources (compared to the 1048 of the \textit{Pantheon} data set) gathered from 18 surveys in the redshift range between $z=0.001$ and $z=2.26$.

The BAO data set here considered is the one of 26 points between $z=0.106$ \citep{2011MNRAS.416.3017B} and $z=2.36$ \citep{2014JCAP...05..027F} described in \citet{2018arXiv180707323S} with the covariance matrix provided by \citet{2016JCAP...06..023S}. This is the same collection already employed in \citet{Dainotti2022Galax..10...24D}, \citet{Dainotti2022PASJ}, \citet{Dainotti2022MNRAS.tmp.2639D}, \citet{Bargiacchi2023arXiv230307076B}, and \citet{Dainotti2023arXiv230510030D}.

\section{Methods}
\label{methods}

\subsection{The physical quantities}
\label{quantities}

We here define the physical quantities we investigate for each probe separately to perform the fit of our cosmographic model. More precisely, we consider the distance moduli for SNe Ia and GRBs, the luminosities for QSOs, and the distance measure for BAOs. These choices have already been described in detail and justified in \citet{Dainotti2022MNRAS.514.1828D}, \citet{Dainotti2022MNRAS.tmp.2639D}, \citet{Bargiacchi2023arXiv230307076B} and \citet{Dainotti2023arXiv230510030D}.

To standardize QSOs as cosmological probes, we use the X-UV relation
\begin{equation}\label{RL}
\mathrm{log_{10}} \, L_{X} = g_1 \, \mathrm{log_{10}} \, L_{UV} + n_1
\end{equation}
where $L_X$ and $L_{UV}$ are the luminosities (in $\mathrm{erg \, s^{-1} \, Hz^{-1}}$) at 2 keV and 2500 \AA, respectively. 
These luminosities are computed from the observed flux densities $F_{X}$ and $F_{UV}$ (in $\mathrm{erg \, s^{-1} \, cm^{-2} \, Hz^{-1}}$), respectively, using $L_{X,UV}= 4 \pi D_{L}^{2}\ F_{X, UV}$. Indeed the K-correction \citep{2001AJ....121.2879B} is omitted for QSOs since their spectral index $\gamma_Q$ is assumed to be 1 in $K = (1+z)^{\gamma_Q -1}$, leading to $K=1$ \citep{2020A&A...642A.150L}. In addition, we also need to correct $L_X$ and $L_{UV}$ for evolutionary effects and selection biases, as detailed in Sect. \ref{epmethod} and as already shown in \citet{Dainotti2013b}, \citet{dainotti2015b}, \citet{DainottiQSO}, \citet{biasfreeQSO2022}, \citet{Bargiacchi2023arXiv230307076B}, \citet{Dainotti2023arXiv230510030D}, and \citet{DainottiGoldQSOApJ2023}. The parameters $g_1$ and $n_1$, and the intrinsic dispersion $sv_1$ of the RL relation, can be fitted through the Kelly technique \citep{Kelly2007}, which is a Bayesian method based on the Markov Chain Monte Carlo (MCMC) approach that allows us to account for uncertainties on all quantities and also for the intrinsic dispersion of the relation.
Using $\mathrm{log_{10}}L_{UV} =  \mathrm{log_{10}}(4 \, \pi \, D_L^2) + \mathrm{log_{10}}F_{UV}$ in Eq. \eqref{RL}, we obtain the observed physical quantity $\mathrm{log_{10}}L_{X,obs}$ under the assumption of the RL relation:
\begin{equation}
\label{lxobs}
\mathrm{log_{10}} \, L_{X,obs} = g_1 \left[ \mathrm{log_{10}}(4 \, \pi \, D_L^2) + \mathrm{log_{10}}F_{UV} \right] + n_1.
\end{equation}
The theoretical quantity is derived as $\mathrm{log_{10}}L_{X,th} = \mathrm{log_{10}}(4 \, \pi \, D_L^2) + \mathrm{log_{10}}F_X$. In both $\mathrm{log_{10}}L_{X,obs}$ and $\mathrm{log_{10}}L_{X,th}$, the luminosity distance $D_L$ (in Megaparsec, Mpc) is given by Eq. \eqref{Dlog} and thus depends on the cosmographic parameters, the redshift, and $H_0$.

To employ GRBs as cosmological probes, we apply the 3D X-ray fundamental plane relation which reads as:
\begin{equation}
\log_{10} L_{\chi} =  a \, \log_{10} T^{*}_{\chi} + b \, \log_{10} L_{peak} + c_1
\label{3drelation}
\end{equation}
where the luminosities $L$ are obtained from the measured fluxes $F$ through the relation $L$, $L= 4 \pi D_L^{2}\, \cdot  F \cdot  \, K$, where $D_L$ is expressed in units of cm and $K$ is the K-correction. In the case of GRBs, since their spectrum can be reproduced with a simple power-law, this correction can be written in terms of the spectral index of the X-ray plateau, $\gamma$, as $K = (1+z)^{\gamma -1}$.
As for QSOs, the free parameters $a$, $b$, and $c_1$ can be evaluated by fitting the 3D relation with the Kelly technique along with the intrinsic dispersion $sv$ of the relation. 
The quantities $L_{\chi}$, $T^{*}_{\chi}$, and $L_{peak}$ in Eq. \eqref{3drelation} should also be corrected for their redshift evolution \citep{Dainotti2022MNRAS.tmp.2639D,Bargiacchi2023arXiv230307076B,Dainotti2023arXiv230510030D}, as detailed in Sect. \ref{epmethod}.
Assuming the 3D relation, we can derive \citep{Dainotti2022MNRAS.tmp.2639D} the observed $\mu$ for GRBs, $\mu_{obs, \mathrm{GRBs}}$, which represents the physical quantity of our interest.
Indeed, using the above-defined relation between luminosities and fluxes to convert luminosities into fluxes in Eq. \eqref{3drelation} and imposing $\mu = 5 \, \mathrm{log_{10}} \, D_L + 25$, we obtain 
\begin{small}
\begin{equation}
\begin{split}
\label{dmGRBs}
\mu_{obs, \mathrm{GRBs}} = & 5 \Bigg[ -\frac{\log_{10} F_{\chi}}{2 (1-b)}+\frac{b \cdot \log_{10} F_{peak}}{2 (1-b)} - \frac{(1-b)\log_{10}(4\pi)+c_1}{2 (1-b)}+ \\ 
& + \frac{a \log_{10} T^{*}_{\chi}}{2 (1-b)} \Bigg] + 25
\end{split}
\end{equation}
\end{small}
where the K-correction has already been applied to all quantities. In this equation, we fix $c_1=23$, as derived in \citet{Dainotti2022MNRAS.tmp.2639D}.
The theoretical $\mu$ is defined as
\begin{equation}
\label{dm}
\mu_{th} = 5 \, \mathrm{log_{10}} \, D_L + 25
\end{equation}
where $D_L$ is given by Eq. \eqref{Dlog}.

For SNe Ia, we use the distance modulus $\mu$ defined as $\mu = m - M$, with $m$ and $M$ the apparent and absolute magnitude, respectively. Since $M$ is degenerate with $H_0$, we cannot determine $H_0$ by using SNe Ia alone \citep{1998A&A...331..815T,scolnic2018}, but $H_0$ can be derived by fixing $M$. In this regard, we directly apply the distance moduli observed, $\mu_{obs, \mathrm{SNe \, Ia}}$, reported by the \textit{Pantheon+} release\footnote{\url{https://github.com/PantheonPlusSH0ES}}, which are computed by fixing $M$ to $M = -19.253 \pm 0.027$, as obtained in \citet{2022ApJ...934L...7R} considering 42 SNe Ia combined with Cepheids hosted in the same galaxies of these SNe Ia. Due to their degeneracy, this assumption on $M$ implies $H_0 = 73.04 \pm 1.04  \, \mathrm{km} \, \mathrm{s}^{-1} \, \mathrm{Mpc}^{-1}$ \citet{2022ApJ...934L...7R}. 
The theoretical $\mu$ is the same defined in Eq. \eqref{dm}.

Regarding BAOs, the considered physical quantity is $d_{z} = r_s(z_d)/D_V(z)$ \citep[e.g.][]{2005ApJ...633..560E}, where $r_s(z_d)$ is the sound horizon at the baryon drag epoch $z_d$ and $D_V(z)$ is the volume averaged distance. We use the observed $d_{z,obs}$ of \citet{2018arXiv180707323S} which are computed from the measured $D_V(z)$ by assuming the fiducial value $(r_s(z_d) \, \cdot h)_{\mathrm{fid}} = 104.57$ Mpc, where $h$ is the dimensionless Hubble constant $\displaystyle h= {H_{0}}/{100 \, \mathrm{km \,s^{-1} \, Mpc^{-1}}}$
\citep[see][]{2016JCAP...06..023S}.
To estimate the theoretical $d_{z,th}$ we use the following numerical approximation \citep{2015PhRvD..92l3516A,2019JCAP...10..044C}, since the exact calculation for $r_s(z_d)$ would require the application of Boltzmann codes:

\begin{equation}
\label{rs}
r_s(z_d) \sim \frac{55.154 \, e^{-72.3  (\Omega_{\nu}\, h^2 + 0.0006)^{2}}}{(\Omega_M \, h^2)^{0.25351} \, (\Omega_b \, h^2)^{0.12807}} \, \mathrm{Mpc}
\end{equation}
where $\Omega_M$, $\Omega_{b}$ and $\Omega_{\nu}$ are the current matter, baryon and neutrino density parameters. In this formula, we fix $\Omega_{\nu} \, h^2 = 0.00064$ and $\Omega_{b} \, h^2 = 0.002237 $ according to \citet{Hinshaw_2013} and \citet{planck2018}).
The theoretical distance $D_V(z)$ required to compute $d_{z,th}$ is defined as \cite{2005ApJ...633..560E}
\begin{equation} \label{DVBAO}
D_{V}(z) = \left[ \frac{cz}{H(z)} \frac{D_L^{2}(z)}{(1+z)^{2}} \right]^{\frac{1}{3}}.
\end{equation}

\subsection{Treatment of redshift evolution and selection biases}
\label{epmethod}

Since GRBs and QSOs are observed up to high-redshifts, we need to correct their observed quantities for possible selection biases and evolutionary effects, which could distort or induce a correlation between physical quantities, thus inducing an incorrect determination of parameters \citep{Dainotti2013a}. To applythese corrections, we employ the Efron \& Petrosian (EP) statistical method \citep{1992ApJ...399..345E} already used in several works for GRBs \citep{Dainotti2013a,dainotti2015b,Dainotti2017A&A...600A..98D,Dainotti2021Galax...9...95D,Dainotti2022MNRAS.tmp.2639D,Bargiacchi2023arXiv230307076B,Dainotti2023arXiv230510030D} and in the QSO realm \citep{DainottiQSO,biasfreeQSO2022,Bargiacchi2023arXiv230307076B,Dainotti2023arXiv230510030D,DainottiGoldQSOApJ2023}. In our study, we use the results of \citet{Dainotti2023arXiv230510030D} for the platinum sample of GRBs and \citet{DainottiQSO} for QSOs and we here just summarize their procedure and outcomes.

In this method, the investigated physical quantities (i.e. luminosities and time) are assumed to evolve with $z$ according to $L' = \frac{L}{(1+z)^{k}}$ and $T' = \frac{T}{(1+z)^{k}}$, where $L$ and $T$ are the observed quantities, $L'$ and $T'$ the corresponding corrected ones without evolution, and $k$ the parameter that mimics the evolution. Nonetheless, the choice of the functional form as a power-law does not affect the results \citep{2011ApJ...743..104S,Dainotti2021ApJ...914L..40D, DainottiQSO} and hence we could also parameterize the dependence on the redshift through more complex functions. 
Then, the the Kendall's $\tau$ statistic is applied to identify the $k$ value that eliminates the evolution with the redshift.
In this procedure, $\tau$ is defined as
\begin{equation}
\label{tau}
    \tau =\frac{\sum_{i}{(\mathcal{R}_i-\mathcal{E}_i)}}{\sqrt{\sum_i{\mathcal{V}_i}}}.
\end{equation}
where $i$ flows along the sources that at $z_i$ present a luminosity greater than the minimum observable luminosity ($L_{min,i}$) at that redshift. $L_{min,i}$ is computed by requiring a limiting flux. The value of this flux threshold is chosen such that the retained sample is composed at least of 90\% of the total initial sources and that it resembles the overall original
distribution according to the Kolmogorov-Smirnov test \citep{Dainotti2013a,dainotti2015b,Dainotti2017A&A...600A..98D,Levine2022ApJ...925...15L,DainottiQSO,Dainotti2022MNRAS.514.1828D}. $\mathcal{R}_i$ is the rank defined as the number of points in the associated set of the $i$-source, where the associated set consists of all $j$-points for which $z_j \leq z_i$ and $L_{z_j} \geq  L_{min,i}$. $\mathcal{R}_i$ is thus computed for each point by considering the position of the source in the sample and including all the sources that can be detected for particular observational limits. In this way the Kendall's statistics is able to take into account and correct for the presence of selection biases in the data, such as the flux limit and temporal resolution of the instrument \citep[see e.g.][for further details]{1992ApJ...399..345E,2006ApJ...642..371K,2009arXiv0909.5051P,Dainotti2013a,Dainotti2013b,DainottiQSO}. In Eq. \eqref{tau}, $\mathcal{E}_i = \frac{1}{2}(i+1)$ and $\mathcal{V}_i = \frac{1}{12}(i^{2}+1)$ are the expectation value and variance, respectively, when the evolution with redshift has been removed. As a consequence, the correlation with redshift disappears when $\tau = 0$, which allows us to obtain the value of $k$ that removes the dependence. The condition $| \tau | > n$ implies that the hypothesis of uncorrelation is rejected at $n \sigma$ level, hence it provides us with the 1 $\sigma$ uncertainty on the $k$ value by imposing $|\tau| \leq 1$. The found value of $k$ can now be used to determine $L'$ (and straightforwardly $T'$ by replacing the luminosity with the time in this procedure) for the total sample.
The $k$ values derived through this procedure for luminosities and time that we here apply are: $k_{L_{peak}} = 1.37^{+0.83}_{-0.93}$, $k_{T_{\chi}} = - 0.68^{+0.54}_{-0.82}$, and $k_{L_{\chi}} = 0.44^{+1.37}_{-1.76}$ for GRBs \citep{Dainotti2023arXiv230510030D}, and $k_{UV} = 4.36 \pm 0.08$ and $k_X = 3.36 \pm 0.07$ for QSOs \citep{DainottiQSO}. The de-evolved quantities computed with these $k$ values are the ones used in our fits when accounting for the redshift evolution. 

Nevertheless, from the description of the EP method, is clear that $k$ is obtained assuming a specific cosmological model, needed to compute the luminosities from the fluxes. We refer to this approach, in which the evolution is taken into account but assuming a specific cosmological model, as ``fixed" correction. In both the works we are referring to \citep[i.e.][]{DainottiQSO,Dainotti2023arXiv230510030D}, the assumed model is a flat $\Lambda$CDM model with $\Omega_M =0.3$ and $H_0 = 70  \, \mathrm{km} \, \mathrm{s}^{-1} \, \mathrm{Mpc}^{-1}$. 
This induces the so-called ``circularity problem".
This problem has been completely overcome for the first time by \citet{Dainotti2022MNRAS.tmp.2639D} for GRBs, and \citet{DainottiQSO} and \citet{biasfreeQSO2022} for QSOs, which have analyzed the trend of $k$ as a function of the cosmological model assumed a-priori. More precisely, in these studies, $k$ is determined not by fixing the cosmological parameters of the assumed model, but over a grid of values of the cosmological parameters (i.e. $\Omega_M$, $H_0$, and also others for models different from the flat $\Lambda$CDM one), leading to the determination of the functions $k(\Omega_M)$ and $k(H_0)$. Even if $k$ does not depend on $H_0$, it shows a dependence on $\Omega_M$ and thus $k(\Omega_M)$ can be applied in the cosmological fits leaving $k$ varies along with the free cosmological parameters, without fixing any cosmology a-priori. We have here described also this method for accounting for the redshift evolution since it is the most appropriate one to avoid the circularity problem. However, in this work we cannot apply this cosmology-independent correction, but only the ``fixed" correction, in which we assume a specific cosmological model. Indeed, we are not dealing with cosmological fits in which cosmological parameters are free to vary and hence we can let $k$ vary with them, but we are investigating a cosmographic model, in which the only free cosmological parameter is $H_0$, which does not affect $k$. Nevertheless, to provide a complete and comprehensive picture of our analysis, we consider four possible treatments of the redshift evolution: without correction and with ``fixed" correction by assuming three different values of $\Omega_M$, $\Omega_M = 0.3$, $\Omega_M = 0.1$, and $\Omega_M = 0.9$ (see Tables \ref{tab:bestfit_QSO} and \ref{tab:bestfit_all}). The investigated value of 0.3 is justified as it is the one expected for the current matter density in the Universe from the latest observations \citep{scolnic2018,2022ApJ...938..110B}. On the other hand, the values of $\Omega_M = 0.1$ and $\Omega_M = 0.9$ have been chosen as illustrative examples corresponding to the highest and lowest values, respectively, that $k$ can assume if $\Omega_M$ ranges in the physically accepted interval between 0.1 and 0.9. Indeed, the function $k(\Omega_M)$ decreases with increasing $\Omega_M$, as shown in Figure 5 of \citet{Dainotti2022MNRAS.tmp.2639D} for GRBs and in Figure 4 of \citet{DainottiQSO} for QSOs. 
Thus, as anticipated above, we assume $k$ fixed according to the chosen $\Omega_M$ value.
We here stress that the reason for investigating these three cases of ``fixed" correction is that, since we cannot apply the above-described technique for the correction for redshift evolution which is model-independent, we test how our results depend on the a-priori assumption of a cosmology. This provides an insight of the impact of our cosmological assumptions which is crucial to interpret the results and draw conclusions.

\subsection{Cosmographic model}
\label{cosmographicmodel}

We here employ the cosmographic model firstly proposed by \citet{2021A&A...649A..65B}, which is an implementation of the one already presented in \citet{rl19} and \citet{lusso2019}. It is an orthogonal polynomial expansion of the luminosity distance in terms of the logarithmic quantity $\log _{10}(1+z)$, where the notation ``orthogonal'' refers to the fact that the coefficients of this power series are not correlated among each other. The uncorrelation between the coefficients of different orders is the main strength of this cosmographic approach since it implies that changing the maximum order of the polynomial does not modify the values of the coefficients of the retained orders. For example, if we move from a fourth-order polynomial to a fifth-order polynomial, the values of the coefficients of the first, second, third, and fourth orders would remain the same compared to the ones of the fourth-order expression. This also implies that, since each value of the coefficients is not correlated to the values of the other coefficients, we can check if an additional higher order in the expansion significantly deviates from zero and thus is needed in the polynomial. Moreover, to quantify the tension between the best-fit of the cosmographic model and the prediction of the flat $\Lambda$CDM model, this feature of orthogonality allows us to determine this discrepancy between the parameters we would obtain from the cosmographic approach and the ones expected from the cosmological model. This can be achieved without the need for accounting for the covariance among the free parameters of the cosmographic model.

As anticipated, the functional form of our cosmographic model simply allows us to determine which is the maximum order of the expansion needed in our study, the latter depending on the data set used. This can be clearly explained if we write down the form of the orthogonal logarithmic polynomial. If we consider a fifth-order expansion, the luminosity distance can be written as\footnote{For sake of simplicity we use $\text{ln}$ instead of $\text{log}_{e}$.}:
\begin{equation}\label{Dlog}
\begin{small}
\begin{split}
&D_{L}(z) =  \frac{c \, \text{ln}(10)}{H_{0}}  \Bigg\{ \log_{10}(1+z) + a_{2} \log_{10}^{2}(1+z) + \\& + a_{3} \left.\Bigg[ k_{32} \log_{10}^{2}(1+z) + \log_{10}^{3}(1+z) \right.\Bigg] + \\& + a_{4} \left.\Bigg[ k_{42} \log_{10}^{2}(1+z) + k_{43} \log_{10}^{3}(1+z) +\log_{10}^{4}(1+z) \right.\Bigg] + \\& + a_{5} \left.\Bigg[ k_{52} \log_{10}^{2}(1+z) + k_{53} \log_{10}^{3}(1+z) + k_{54} \log_{10}^{4}(1+z) + \log_{10}^{5}(1+z) \right.\Bigg] \Bigg\} .
\end{split}
\end{small}
\end{equation}
The first term, $\displaystyle \frac{c \, \text{ln}(10)}{H_{0}}$, where $c$ is the speed of light, and $a_{1}=1$ are imposed to recover the Hubble law at low $z$, which is $\displaystyle D_{L}(z) = \frac{c \, \text{ln}(1+z)}{H_{0}}$. The coefficients $k_{ij}$ are not free parameters, and depend on the probes considered and, especially, on their redshift distribution. Indeed, their values are determined through the following step-by-step procedure such that all the $a_{i}$ in Eq. \eqref{Dlog} are uncorrelated (for visualization see Figs. \ref{fig: QSO_Gauss}, \ref{fig: QSO_New}, \ref{fig: all_Gauss}, and \ref{fig: all_New}). This technique is completely general, it can be applied to any probe, and it should be performed any time the sample investigated changes.

As a first step, the considered data set is fitted with a second-order non-orthogonal polynomial $\displaystyle P_{2} = \frac{c \, \text{ln}(10)}{H_{0}}\Bigg[\log_{10}(1+z) + a'_{2}\log_{10}^{2}(1+z)\Bigg]$ to obtain $a'_{2}$. Then, the fit is performed with a third-order non-orthogonal polynomial $\displaystyle P_{3} = \frac{c \, \text{ln}(10)}{H_{0}}\Bigg[\log_{10}(1+z) + a''_{2}\log_{10}^{2}(1+z) + a''_{3}\log_{10}^{3}(1+z)\Bigg]$ to determine $a''_{2}$ and $a''_{3}$. By comparing the orthogonal and non-orthogonal polynomial expressions we obtain the constraint $ a''_{2}=a'_{2}+a''_{3} k_{32}$. 
We apply the same procedure to the other order polynomials.
Moving forward with the fourth-order non-orthogonal polynomial $\displaystyle P_{4} = \frac{c \, \text{ln}(10)}{H_{0}}\Bigg[\log_{10}(1+z) + a'''_{2}\log_{10}^{2}(1+z) + a'''_{3}\log_{10}^{3}(1+z) + a'''_{4}\log_{10}^{4}(1+z)\Bigg]$ we obtain $a'''_{2}$, $a'''_{3}$, and $a'''_{4}$, and with the fifth-order non-orthogonal polynomial $\displaystyle P_{5} = \frac{c \, \text{ln}(10)}{H_{0}}\Bigg[\log_{10}(1+z) + a''''_{2}\log_{10}^{2}(1+z) + a''''_{3}\log_{10}^{3}(1+z) + a''''_{4}\log_{10}^{4}(1+z) + a''''_{5}\log_{10}^{5}(1+z)\Bigg]$ we determine $a''''_{2}$, $a''''_{3}$, $a''''_{4}$, and $a''''_{5}$, Hence we impose the following equivalences:
\begin{align}
& a'''_{2}=a'_{2}+a''_{3}k_{32}+a'''_{4}k_{42},\\
& a'''_{3}=a''_{3}+a'''_{4}k_{43},\\
& a''''_{2}=a'_{2}+a''_{3}k_{32}+a'''_{4}k_{42}+a''''_{5}k_{52},\\
& a''''_{3}=a''_{3}+a'''_{4}k_{43}+a''''_{5}k_{53},\\
& a''''_{4}=a'''_{4}+a''''_{5}k_{54}.
\end{align}
The solutions to this system of equations are the six unknown coefficients $k_{ij}$. The same procedure can be straight-forwardly employed to determine $k_{ij}$ for the logarithmic orthogonal polynomial of any order (greater or lower than the fifth order). We refer to \citet{2021A&A...649A..65B} for further detail on the overall procedure.

From Eq. \eqref{Dlog} it is clearly visible that the higher-order coefficients become significant only at high redshifts. It means that the higher is the maximum redshift reached by the probes in our data set and the larger is the number of sources at high redshifts in our sample, and the higher is the maximum order of the polynomial needed to fit the cosmographic model. As a consequence, in our study, we have separately considered the two data sets used, QSOs alone and QSOs combined with GRBs, SNe Ia, and BAOs, to identify the proper polynomial expansion to be employed in the two different cases. As anticipated in Section \ref{intro}, the flexibility of this cosmographic approach, that allows us to choose the most suitable functional form for the data investigated, guarantees an accurate fit of our data.
We here also stress that the cosmographic approach we employ has already been validated in \citet{2021A&A...649A..65B} against possible issues such as convergence problems when the cosmographic function is fitted over a redshift range wider than its convergence radius, or troubles in the comparison between a cosmographic fit on the whole redshift range and the theoretical prediction of a given physical model, which is instead computed around $z = 0$ (see also Section \ref{comparison}).

\subsection{Comparison with a flat $    \boldsymbol{\Lambda}$CDM model}
\label{comparison}

As anticipated in Section \ref{intro} and  detailed in \citet{2021A&A...649A..65B}, to quantify the discrepancy between the cosmographic fit and a flat $\Lambda$CDM model, we can compare the best-fit values of the $a_i$ coefficients in Eq. \eqref{Dlog} (or for any order of the orthogonal polynomial logarithmic expansion of $D_L(z)$) to the one predicted in a flat $\Lambda$CDM model. To this end, we need to expand in power series of $z$ both Eq.\eqref{Dlog} and the expression of  $D_L(z)$ in a flat $\Lambda$CDM model, which is
\begin{equation}
\label{Dl_lcdm}
D_{L, \mathrm{flat \Lambda \, CDM}}(z) = (1+z) \frac{c}{H_{0}} \, \int^{z}_{0} \frac{d z'}{\sqrt{\Omega_{M} (1+z)^{3} + (1- \Omega_{M})}},
\end{equation}. Then, comparing the coefficients of the corresponding orders of the two obtained expansions, we obtain the required equations that link the cosmographic parameters to the cosmological ones.
The relations between the free parameters $a_{i}$ of the fourth-order cosmographic logarithmic expansion and $\Omega_M$ in a flat \text{$\Lambda$}CDM model are the following:
\begin{subequations}
\label{coefflcdm_4}
\begin{align}
& a_{4}=\text{ln}^{3}(10)\left(-\frac{135}{64} \Omega^{3}_{M} + \frac{18}{4} \Omega^{2}_{M} - \frac{47}{16} \Omega_M +\frac{5}{8}\right),\\
& a_{3}=-k_{43}a_{4} + \text{ln}^{2}(10) \left( \frac{9}{8} \Omega^{2}_{M} - 2 \Omega_M + \frac{7}{6} \right),\\
& a_{2}=-k_{32}a_{3}-k_{42}a_{4}+\text{ln}(10) \left( -\frac{3}{4} \Omega_M + \frac{3}{2} \right) .
\end{align}
\end{subequations}
Extending to the fifth-order we obtain:
\begin{subequations}
\label{coefflcdm_5}
\begin{align} 
& a_{5}=\text{ln}^{4}(10) \left( \frac{567}{128} \Omega^{4}_{M}-\frac{729}{64}\Omega^{3}_{M}+\frac{315}{32}\Omega^{2}_{M}-\frac{25}{8}\Omega_M+\frac{31}{120} \right),\\
& a_{4}=-k_{54}a_{5}+\text{ln}^{3}(10)\left(-\frac{135}{64} \Omega^{3}_{M} + \frac{18}{4} \Omega^{2}_{M} - \frac{47}{16} \Omega_M +\frac{5}{8}\right),\\
& a_{3}=-k_{53}a_{5}-k_{43}a_{4} + \text{ln}^{2}(10) \left( \frac{9}{8} \Omega^{2}_{M} - 2 \Omega_M + \frac{7}{6} \right),\\
& a_{2}=-k_{32}a_{3}-k_{42}a_{4}-k_{52}a_{5}+\text{ln}(10) \left( -\frac{3}{4} \Omega_M + \frac{3}{2} \right) .
\end{align}
\end{subequations}
Tables \ref{tab:bestfit_QSO} and \ref{tab:lcdm_QSO} show, respectively, the best-fit $a_i$ values and the predicted $a_i$ values in a flat $\Lambda$CDM model under the different assumptions considered in this work when only QSOs are considered. Tables \ref{tab:bestfit_all} and \ref{tab:lcdm_all} report the same, but for all probes together.

\subsection{Likelihoods and fitting procedure}
\label{fittingprocedure}

The cosmographic fits presented in this work are performed with the Kelly method \citep{Kelly2007} in which all
parameters are free to vary contemporaneously. Indeed, when considering QSOs alone (see Figs. \ref{fig: QSO_Gauss} and \ref{fig: QSO_New}), we fit the parameters of the RL relation, $g_1$, $n_1$, and $sv_1$, the coefficients $a_i$ of the cosmographic model, and $H_0$, which is included in the cosmographic expansion of Eq. \eqref{Dlog}. Instead, when joining QSOs, GRBs, SNe Ia, and BAOs, we deal with the additional free parameters of the 3D relation, $a$, $b$, and $sv$ (see Figs. \ref{fig: all_Gauss} and \ref{fig: all_New}). In both cases, we impose wide uniform priors on all free parameters.
As already discussed in \citet{Dainotti2023arXiv230510030D}, to guarantee the reliability of the posterior distributions of the fitted parameters, 
we have employed the Gelman-Rubin convergence diagnostic test, which relies on the square root of the ratio between the variance intra-chain (within each chain) and the variance inter-chain (between one chain and another) ($R$), which is applied as a convergence probe: if $R$ is large
the convergence has not yet been reached. Following \citet{biasfreeQSO2022}, \citet{Bargiacchi2023arXiv230307076B}, and \citet{Dainotti2023arXiv230510030D}, we impose $R-1 < 0.05$ to guarantee the convergence of each free parameter, which is more restrictive than $R-1 < 0.1$ often required in the literature \citep{DainottiQSO}. 

In our analysis, we not only distinguish between our two different data sets (i.e. QSOs alone and the combination of QSOs, GRBs, SNe Ia, and BAOs), but also between the application of different likelihoods. Indeed, \citet{snelikelihood2022_all}, \citet{snelikelihood2022_all}, \citet{Bargiacchi2023arXiv230307076B}, and \citet{Dainotti2023arXiv230510030D} have proved through several statistical tests that the proper best-fit likelihoods for the samples here studied of QSOs, SNe Ia, and BAOs are not the commonly used Gaussian likelihoods, while they are a logistic for QSOs and a student-T for \textit{Pantheon+} SNe Ia and BAOs. Only GRBs from the Platinum sample are properly fitted through a Gaussian likelihood. The above-mentioned works have also shown the importance of choosing the proper likelihood for each probe to significantly reduce the uncertainties on the fitted parameters. We here stress that the newly found distributions do not effectively reveal the truth about underlying likelihoods of QSOs, SNe Ia, and BAOs but they reproduce the actual distributions of these probes better than the commonly used Gaussian distribution. In this sense, the new best-fit distributions represent a useful and more precise parameterization to constrain parameters with increased accuracy, as shown in \citet{Bargiacchi2023arXiv230307076B}, \citet{Dainotti2023arXiv230510030D}, and \citet{snelikelihood2022_all}. 
Hence, in this work we separately consider the cases in which the data sets (QSOs alone or the combination of all four probes) are fitted with Gaussian likelihoods ($\cal L$$_{Gaussian}$) and with the new best-fit proper likelihoods ($\cal L$$_{new}$). When fitting all probes together, we employ the joint likelihood, as discussed in detail in \citet{Dainotti2023arXiv230510030D}, and, in the case of $\cal L$$_{new}$ we consider as an additional free parameter the degree of freedom of the student-T distribution, $\nu_{\mathrm{SNe}}$ and $\nu_{\mathrm{BAO}}$, for SNe Ia and BAOs respectively, as shown in Figure \ref{fig: all_New}.

\subsubsection{The posterior likelihood}

We here detail the posterior likelihood mentioned above. Indeed, the posterior probability is obtained as the sum (or the product if we consider the natural logarithm units) of the prior probability and the likelihood. As stated before, we impose a uniform (i.e. uninformative) prior on all free parameters in the following wide ranges:
$0.1 < g_1 < 1$, $2< n_1 < 20$, $0 < sv_1 < 2$, $50 \, \mathrm{km} \, \mathrm{s}^{-1} \, \mathrm{Mpc}^{-1} \leq H_0 \leq 100 \, \mathrm{km} \, \mathrm{s}^{-1} \, \mathrm{Mpc}^{-1}$, $-100 \leq a_2 \leq 150$, $-100 \leq a_3 \leq 150$, $-100 \leq a_4 \leq 150$, $-100 \leq a_5 \leq 150$, and, when also GRBs, SNe Ia, and BAOs are considered, $0 < \nu_{\mathrm{SNe}} < 10$, $0 < \nu_{\mathrm{BAO}} <10$, $-2 < a < 0$, $0 < b < 2$, and $0< sv < 2$. Concerning the likelihoods, when we investigate $\cal L$$_{Gaussian}$, the likelihood is defined for each individual probe as follows, where ln stands for the natural logarithm and $\mathcal{L}$ for the likelihood. For QSOs,
\begin{equation} \label{lfqso}
\text{ln}(\mathcal{L})_{\text{QSOs}} = -\frac{1}{2} \sum_{i=1}^{N} \left[ \frac{(\mathrm{log_{10}}L_{X,obs,i}-\mathrm{log_{10}}L_{X,th,i})^{2}}{\hat{s}^{2}_{i}} + \text{ln}(\hat{s}^{2}_{i})\right]
\end{equation}
where $\hat{s}^{2}_{i} = \sigma ^{2}_{\mathrm{log_{10}}L_{X,obs,i}} + g_1^{2} \sigma ^{2}_{\mathrm{log_{10}}L_{UV,obs,i}} + sv_1^{2}$ in which the statistical uncertainties on $\mathrm{log_{10}}L_{X,obs}$ and $\mathrm{log_{10}}L_{UV,obs}$ and the intrinsic dispersion $sv_1$ are taken into account.
Similarly, for GRBs,
\begin{equation} \label{lfgrb}
\text{ln}(\mathcal{L})_{\text{GRBs}} = -\frac{1}{2} \sum_{i=1}^{N} \left[ \frac{(\mu_{obs, \mathrm{GRBs},i} -\mu_{th,i})^{2}}{s^{2}_{i}} + \text{ln}(s^{2}_{i})\right]
\end{equation}
where $s^2_{i} = \sigma ^{2}_{\mu_{obs, \mathrm{GRBs}}} + sv^{2}$ which accounts for the statistical uncertainties on the measured quantities but also the intrinsic dispersion $sv$.
For SNe Ia instead $\mathcal{L}$ is built as
\begin{equation} \label{lfsne}
\text{ln}(\mathcal{L})_{\mathrm{SNe Ia}} = -\frac{1}{2} \Bigg[\left(\boldsymbol{\mu_{obs, \mathrm{SNe \, Ia}}}-\boldsymbol{\mu_{th}}\right)^{T} \, \textit{C}^{-1} \, \left(\boldsymbol{\mu_{obs, \mathrm{SNe \, Ia}}}-\boldsymbol{\mu_{th}}\right)\Bigg]
\end{equation}
where $\textit{C}$ is the covariance matrix that includes both statistical and systematic uncertainties on the measured distance moduli provided by \textit{Pantheon +} release.
In the case of BAOs, $\mathcal{L}$ reads as
\begin{equation} 
\label{lfbao2}
\text{ln}\left(\mathcal{L}\right)_{\mathrm{BAO}} = -\frac{1}{2} \left[\left(d_{z,obs}-d_{z,th}\right)^{T}\, \mathcal{C}^{-1}\, \left(d_{z,obs}-d_{z,th}\right)\right]
\end{equation}
with $\mathcal{C}$ the covariance matrix provided in \citet{2016JCAP...06..023S}.\\
On the other hand, when applying the new best-fit likelihoods $\cal L$$_{new}$, the formula for GRBs remains unchanged (Eq. \eqref{lfgrb}) since the likelihood is still Gaussian, while for the other probes we need to employ the actual probability distribution functions (PDFs). Hence, in the case of QSOs $\mathcal{L}$ follows from a logistic PDF and reads as:
\begin{equation} \label{lfqso_new}
\text{ln}(\mathcal{L})_{\text{QSOs}} = \text{ln} ( \mathrm{PDF_{logistic}}) = \text{ln} \frac{e^{-\frac{(\mathrm{log_{10}}L_{X,obs}-\mathrm{log_{10}}L_{X,th})}{\xi}}}{\xi \, \left(1+ e^{\frac{-(\mathrm{log_{10}}L_{X,obs}-\mathrm{log_{10}}L_{X,th})}{\xi}}\right)^2},
\end{equation}
where $\xi$ is the scale parameter for which $\hat{s}^{2} = (\xi^2 \, \pi^2)/3$.
Similarly, we replace Eqs. \eqref{lfsne} and \eqref{lfbao2} with the likelihoods obtained from a student-T PDF. Thus, for SNe Ia 
\begin{equation} \label{lfsne_new}
\begin{split}
& \text{ln}(\mathcal{L})_{\mathrm{SNe Ia}}  = \text{ln} (\mathrm{PDF_{student}})  = \\ & =  \text{ln} \frac{\Gamma\left(\frac{\nu_{\mathrm{SNe}} +1}{2}\right)}{\sqrt{\nu_{\mathrm{SNe}} \, \pi} \, \xi \, \Gamma \left(\frac{\nu_{\mathrm{SNe}}}{2}\right)} \, \left[1 + \frac{((\mu_{obs, \mathrm{SNe \, Ia}}-\mu_{th})/\xi)^2}{\nu_{\mathrm{SNe}}}\right]^{-\frac{\nu_{\mathrm{SNe}} +1}{2}}\,,
\end{split}
\end{equation}
where $\Gamma$ is the $\gamma$ function and the variance is $\sigma^2 = (\xi^2 \, \nu_{\mathrm{SNe}}) / (\nu_{\mathrm{SNe}} -2)$.
Instead, considering BAOs:
\begin{equation} \label{lfbao_new}
\begin{split}
& \text{ln}(\mathcal{L})_{\mathrm{BAO}}  = \text{ln} (\mathrm{PDF_{student}})  = \\ & =  \text{ln} \frac{\Gamma\left(\frac{\nu_{\mathrm{BAO}} +1}{2}\right)}{\sqrt{\nu_{\mathrm{BAO}} \, \pi} \, \xi \, \Gamma \left(\frac{\nu_{\mathrm{BAO}}}{2}\right)} \, \left[1 + \frac{((d_{z,obs}-d_{z,th})/\xi)^2}{\nu_{\mathrm{BAO}}}\right]^{-\frac{\nu_{\mathrm{BAO}} +1}{2}}\,,
\end{split}
\end{equation}
where the variance is $\sigma^2 = (\xi^2 \, \nu_{\mathrm{BAO}}) / (\nu_{\mathrm{BAO}} -2)$.\\
We here stress that, as anticipated in Sect. \ref{fittingprocedure}, when combining all the probes, we use the joint likelihood, that is the sum of all the $\text{ln} \cal{L}$ above-defined \citep[see][for details]{Dainotti2023arXiv230510030D}. This joint likelihood (or only the one of QSOs when QSOs alone are used) is then added to the natural logarithm of the prior probability to obtain the posterior likelihood.

\section{Results and Discussion}
\label{results}

\subsection{Results from QSOs alone}
\label{results_QSO}

\begin{figure*}
\centering
\begin{subfigure}[b]{.45\linewidth}
\includegraphics[width=\linewidth]{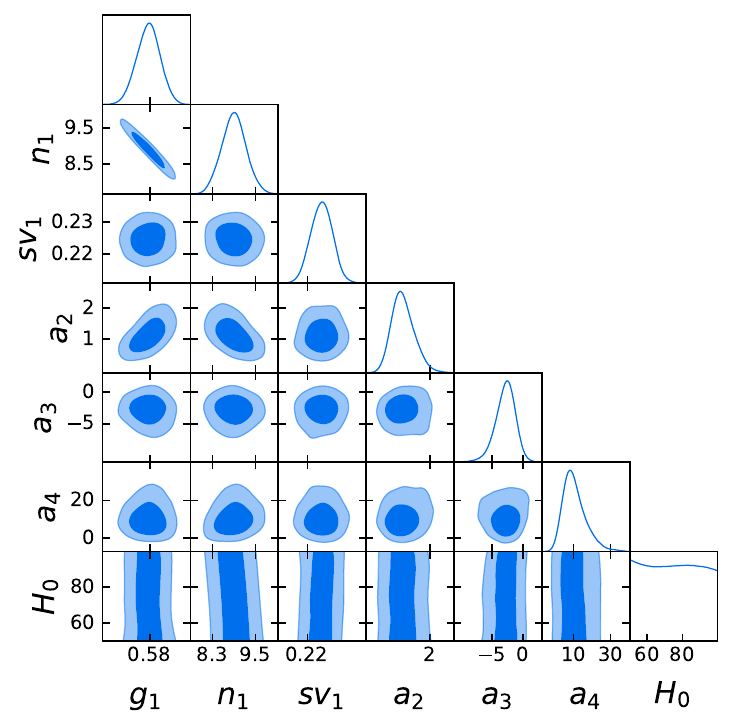}
\caption{Without correction for evolution}\label{}
\end{subfigure}
\begin{subfigure}[b]{.45\linewidth}
\includegraphics[width=\linewidth]{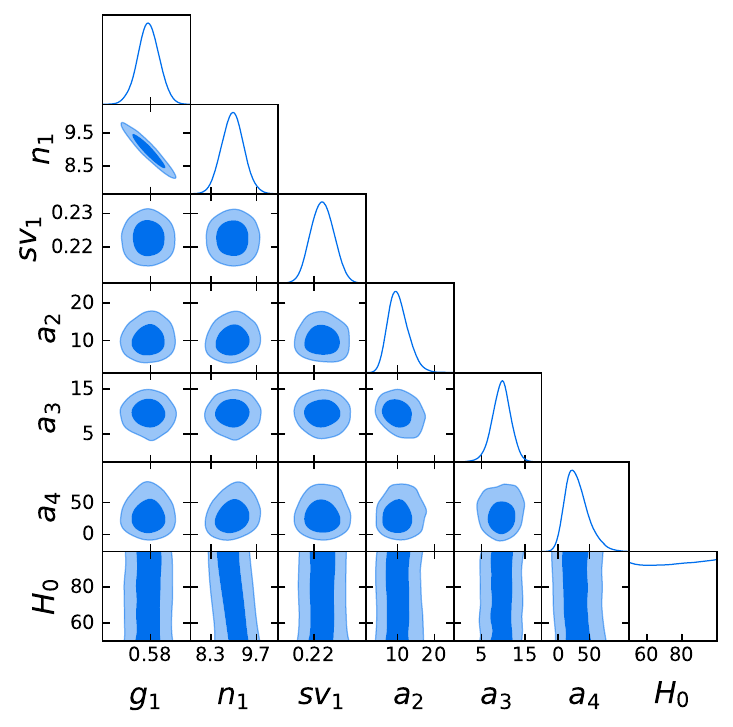}
\caption{With fixed correction for evolution assuming $\Omega_M = 0.3$}\label{}
\end{subfigure}
\caption{Fit of QSOs alone with $\cal L$$_{Gaussian}$ for the evolutionary cases without correction for evolution (left panel) and with fixed correction for evolution assuming $\Omega_M=0.3$ (right panel). For sake of visualization clarity, we do not show also the other two cases of fixed correction with $\Omega_M=0.1$ and $\Omega_M=0.9$ investigated in this work. These additional cases present similarities to the ones here shown and are described in detail in Table \ref{tab:bestfit_QSO} and discussed in Sect. \ref{results_QSO}.}
\label{fig: QSO_Gauss}
\end{figure*}

\begin{figure*}
\centering
\begin{subfigure}[b]{.45\linewidth}
\includegraphics[width=\linewidth]{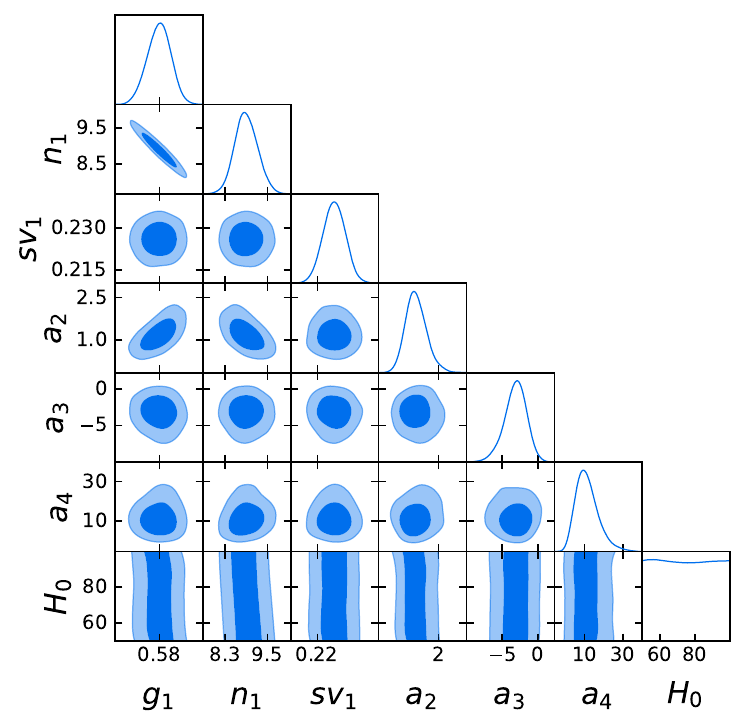}
\caption{Without correction for evolution}\label{}
\end{subfigure}
\begin{subfigure}[b]{.45\linewidth}
\includegraphics[width=\linewidth]{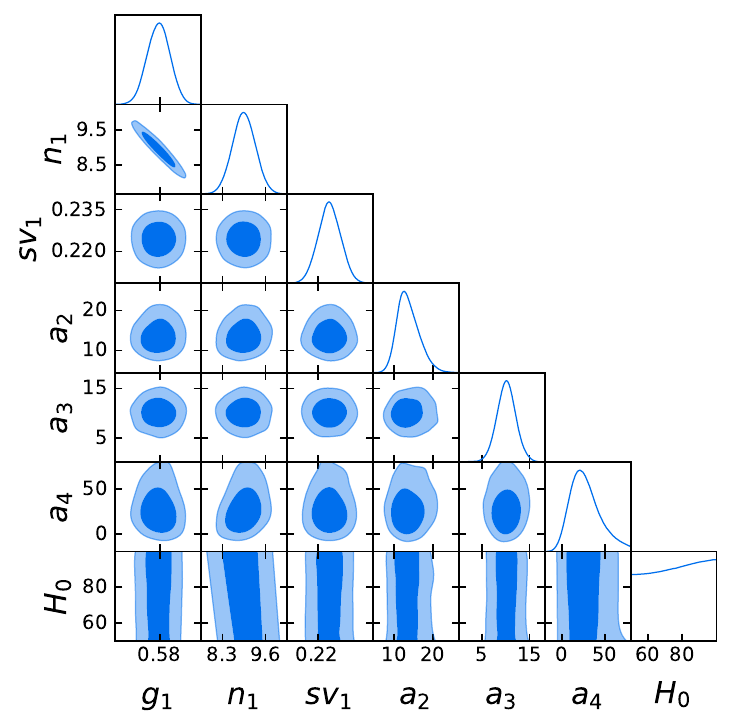}
\caption{With fixed correction for evolution assuming $\Omega_M = 0.3$}\label{}
\end{subfigure}
\caption{Fit of QSOs alone with $\cal L$$_{new}$ for the evolutionary cases without correction for evolution (left panel) and with fixed correction for evolution assuming $\Omega_M=0.3$ (right panel). For sake of visualization clarity, we do not show also the other two cases of fixed correction with $\Omega_M=0.1$ and $\Omega_M=0.9$ investigated in this work. These additional cases present similarities to the ones here shown and are described in detail in Table \ref{tab:bestfit_QSO} and discussed in Sect. \ref{results_QSO}.}
\label{fig: QSO_New}
\end{figure*}

We here outline the results obtained by fitting QSOs alone.
\begin{itemize}
    \item The first step in our cosmographic procedure is to determine the order of the orthogonal logarithmic polynomial that is required to properly fit the data. As detailed in Section \ref{cosmographicmodel}, the maximum order needed depends on the sample considered. When using only QSOs, we obtain that the fifth-order is negligible, since in all cases studied the best-fit value of $a_5$ is consistent with 0 within much less than 1 $\sigma$. Hence, we truncate the formula of Eq. \eqref{Dlog} to the fourth order imposing $a_5=0$. Indeed, as visible from Table \ref{tab:bestfit_QSO} and Figs. \ref{fig: QSO_Gauss} and \ref{fig: QSO_New}, $a_4$ is significant since it is always discrepant from $a_4=0$ at $> 1 \, \sigma$ level.

    \item Comparing the results from the different likelihoods in Table \ref{tab:bestfit_QSO}, we can observe that $\cal L$$_{new}$ leads to reduced uncertainties on the parameters compared to $\cal L$$_{Gaussian}$, as already proved in the cosmological analyses of \citet{snelikelihood2022_all}, \citet{Bargiacchi2023arXiv230307076B}, and \citet{Dainotti2023arXiv230510030D}. More specifically, in our case $\cal L$$_{new}$ in general lowers the errors on the higher-order $a_i$ coefficients, reaching a reduction of $\sim 23 \%$. Nevertheless, QSOs alone cannot yield tight constraints on the coefficients $a_i$, which are indeed associated with large 1 $\sigma$ uncertainties.

    \item Concerning instead the only cosmological parameters, $H_0$, it is not constrained in any of the cases studied with QSOs alone, as shown in Figs. \ref{fig: QSO_Gauss} and \ref{fig: QSO_New}. Indeed, QSOs alone not calibrated on SNe Ia, which behave as anchors for the low-$z$ regime, cannot determine $H_0$, as already discussed in \citet{Dainotti2023arXiv230510030D}.

    \item For both the likelihoods considered, a trend between the three cases (i.e. $\Omega_M=0.3$, $\Omega_M=0.1$, and $\Omega_M=0.9$)  with fixed correction for evolution and the case without correction is exhibited: the application of the correction leads to higher uncertainties of $a_i$ since the evolutionary parameter $k$ brings an additional uncertainty. Moreover, the cases with correction significantly changes the best-fit values of the coefficients compared to the ones obtained without the correction for the evolution.
    We can also notice that the case with correction with $\Omega_M=0.9$ is the one that leads to best-fit values more similar to the ones obtained without evolution. This is completely in agreement with \citet{biasfreeQSO2022}, where it is shown that QSOs alone, without calibration on SNe Ia and without accounting for any correction for the redshift evolution, prefer $\Omega_M$ towards 0.9 (see their Table 2 and Figure 7). Indeed, these two configurations, the one without correction and the one with correction and $\Omega_M=0.9$, are the ones that bring the lowest uncertainties on $a_2$, $a_3$, and $a_4$.

    \item Focusing on the best-fit values of the $a_i$ coefficients, as anticipated, we obtain very discrepant values comparing the different evolutionary cases for each likelihood studied, even though also the uncertainties are large. When comparing instead the best-fit values of the coefficient of the same order for $\cal L$$_{Gaussian}$ and $\cal L$$_{new}$ and same cases of evolutionary correction, we notice that all $a_i$ are consistent within less than 1 $\sigma$, thus the two likelihoods yield compatible best-fit values, in agreement with the results of \citet{Bargiacchi2023arXiv230307076B} and \citet{Dainotti2023arXiv230510030D}.

    \item Through Eqs. \eqref{coefflcdm_4}, we compute the discrepancy between the obtained best-fit values of $a_i$ reported in Table \ref{tab:bestfit_QSO} and the theoretically predicted values in a flat $\Lambda$CDM model (Table \ref{tab:lcdm_QSO}) for the same cases. There is only one case of compatibility within 1 $\sigma$: $a_3$ with correction for redshift evolution with $\Omega_M=0.9$ for both likelihoods. More precisely, in this case we obtain compatibility in 0.3 $\sigma$ and 0.4 $\sigma$, for $\cal L$$_{Gaussian}$ and $\cal L$$_{new}$, respectively. In all the other cases studied, a tension with the flat $\Lambda$CDM model emerges with a significance level $> 1 \, \sigma$, that reaches up to $\sim 14 \, \sigma$ on $a_2$ in the case without correction for redshift evolution for both the likelihoods.
    
\end{itemize}

\subsection{Results from all probes together}
\label{results_all}

\begin{figure*}
\centering
\begin{subfigure}[b]{.45\linewidth}
\includegraphics[width=\linewidth]{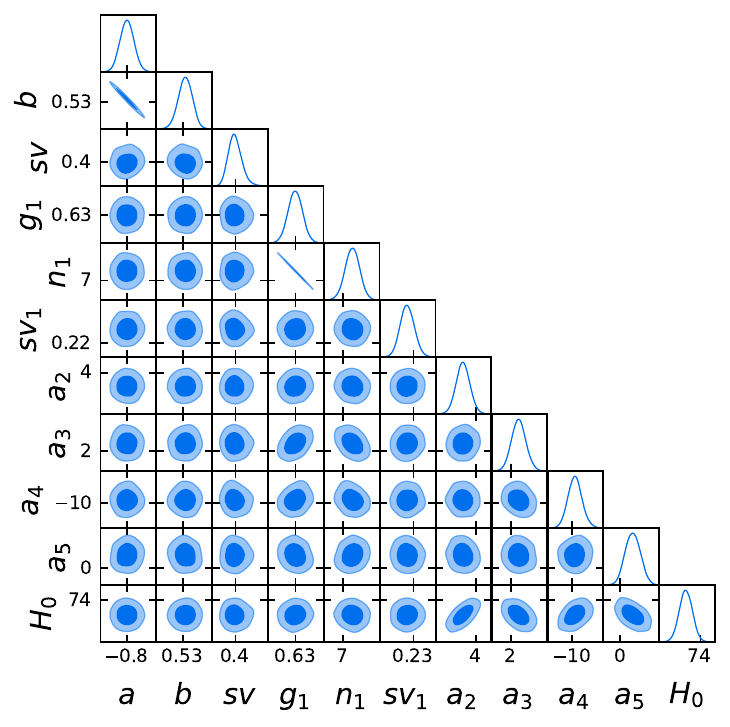}
\caption{Without correction for evolution}\label{}
\end{subfigure}
\begin{subfigure}[b]{.45\linewidth}
\includegraphics[width=\linewidth]{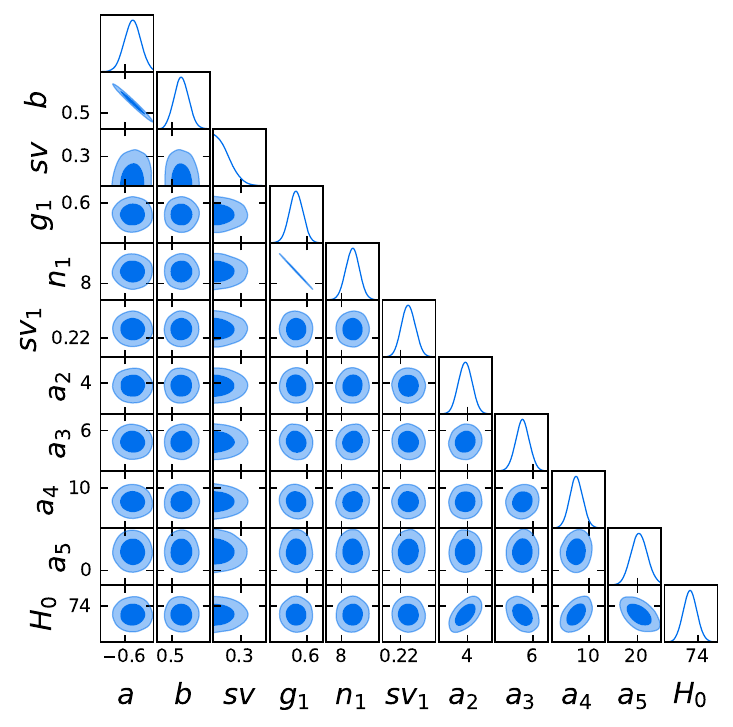}
\caption{With fixed correction for evolution assuming $\Omega_M = 0.3$}\label{}
\end{subfigure}
\caption{Fit of all probes together with $\cal L$$_{Gaussian}$ for the evolutionary cases without correction for evolution (left panel) and with fixed correction for evolution assuming $\Omega_M=0.3$ (right panel). For sake of visualization clarity, we do not show also the other two cases of fixed correction with $\Omega_M=0.1$ and $\Omega_M=0.9$ investigated in this work. These additional cases present similarities to the ones here shown and are described in detail in Table \ref{tab:bestfit_all} and discussed in Sect. \ref{results_all}.}
\label{fig: all_Gauss}
\end{figure*}

\begin{figure*}
\centering
\begin{subfigure}[b]{.45\linewidth}
\includegraphics[width=\linewidth]{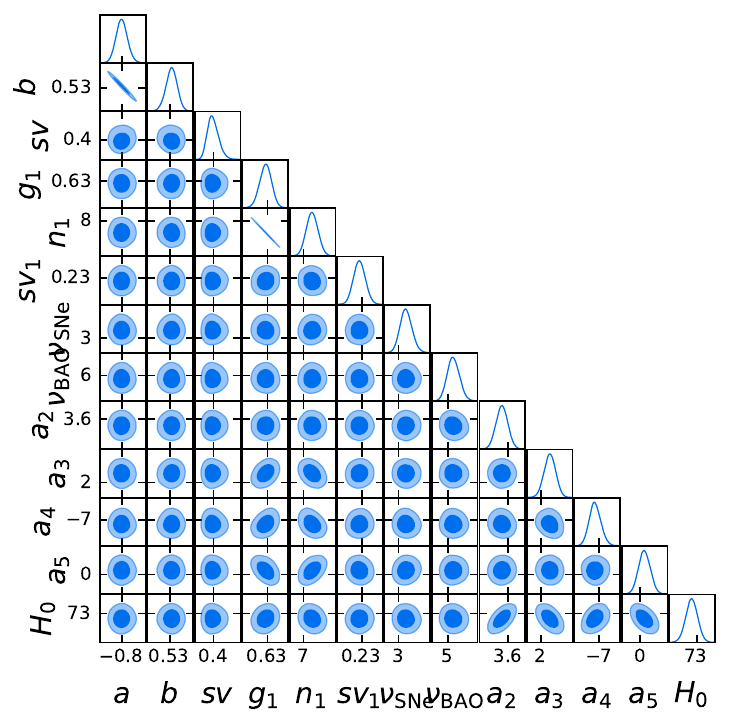}
\caption{Without correction for evolution}\label{}
\end{subfigure}
\begin{subfigure}[b]{.45\linewidth}
\includegraphics[width=\linewidth]{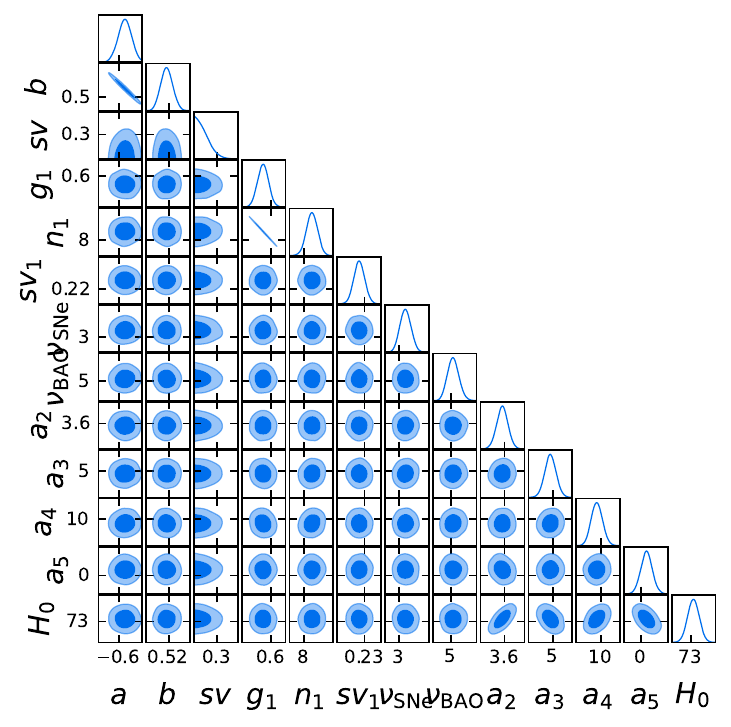}
\caption{With fixed correction for evolution assuming $\Omega_M = 0.3$}\label{}
\end{subfigure}
\caption{Fit of all probes together with $\cal L$$_{new}$ for the evolutionary cases without correction for evolution (left panel) and with fixed correction for evolution assuming $\Omega_M=0.3$ (right panel). For sake of visualization clarity, we do not show also the other two cases of fixed correction with $\Omega_M=0.1$ and $\Omega_M=0.9$ investigated in this work. These additional cases present similarities to the ones here shown and are described in detail in Table \ref{tab:bestfit_all} and discussed in Sect. \ref{results_all}.}
\label{fig: all_New}
\end{figure*}

As already discussed in detail in \citet{2022MNRAS.515.1795B} and \citet{Dainotti2023arXiv230510030D}, a debate has recently emerged on the importance of properly combining data sets that do not reveal manifest tension to infer correct statistical interpretations from the analyses \citep{planck2018,Efstathiou:2020wem,2021ApJ...908...84V,2021JCAP...11..060G}. In this regard, QSOs pose a problem since, as standalone probes, determine parameters with no tight constraints and large uncertainties, as stressed above. Thus, to assess the combination of QSOs with other probes and to verify if all the probes can be joint, in future studies, we would need to strongly reduce the intrinsic dispersion of the RL relation such that QSOs could reach the precision of SNe Ia in constraining parameters as it has been already done in \citet{DainottiGoldQSOApJ2023}. 
The aim here however is to understand to what extend the total sample of QSOs adoped for cosmological analyses can indeed be used in the cosmographic approach. Future studies may entail the use of cosmographic applications with this golden sample. 
Beside the relevance of this topic, it is beyond the aim of our analysis, and we here discuss the results obtained by combining QSOs, GRBs, SNe Ia, and BAOs, analyzing the different cases studied, the difference and similarities with the results from QSOs alone, and comparing with previous works in the literature.
\begin{itemize}
    \item Starting from the maximum order of the orthogonal logarithmic polynomial needed to fit all probes together, we notice from Table \ref{tab:bestfit_all} and Figs. \ref{fig: all_Gauss} and \ref{fig: all_New} that in all the cases with $\cal L$$_{Gaussian}$ $a_5$ is significant with a discrepancy from $a_5=0$ always greater than 1.5 $\sigma$. We have further checked that instead the addition of the sixth-order leads to $a_6$ compatible with $a_6=0$. Hence, in this cases we have used the fifth-order function, as in Eq. \eqref{Dlog}. The discrepancy of $a_5$ from 0 appears to be reduced when using $\cal L$$_{new}$, reaching the minimum of 0.2 $\sigma$ in the case with the correction for evolution applied by assuming $\Omega_M=0.9$. This shows that the use of $\cal L$$_{new}$, in place of $\cal L$$_{Gaussian}$, reduces the dimensionality of the cosmographic polynomial. Nevertheless, we still fit a fifth-order cosmographic model (Eq. \eqref{Dlog}) when applying $\cal L$$_{new}$ to be consistent with the case of $\cal L$$_{Gaussian}$. As already stated for QSOs alone, we can also notice that $\cal L$$_{new}$ reduce the uncertainties on the parameters compared to $\cal L$$_{Gaussian}$, mainly on the higher-order coefficients, up to a maximum reduction of $\sim 20 \%$, and on $H_0$, with an uncertainty reduced of $\sim 28 \%$. This reduction on the error on $H_0$ obtained through $\cal L$$_{new}$ confirms the results of \citet{snelikelihood2022_all} and \citet{Dainotti2023arXiv230510030D}.

    \item Still focusing on $H_0$, we observed that, differently from the case of only QSOs, the combination of the four probes allows us to obtain close contours on $H_0$ in all cases studied. The value of $H_0$ is always around $73 \, \mathrm{km} \, \mathrm{s}^{-1} \, \mathrm{Mpc}^{-1}$ within 0.5 $\sigma$, since it is mainly driven by \textit{Pantheon+} SNe Ia and BAOs \citep[see e.g.][]{2022ApJ...934L...7R,Dainotti2022MNRAS.tmp.2639D,Bargiacchi2023arXiv230307076B,Dainotti2023arXiv230510030D}, and it is in all cases compatible within 0.5 $\sigma$ with the value $H_0=73.04 \pm 1.04$ reported in \citet{2022ApJ...934L...7R}. 
    In addition, the use of $\cal L$$_{new}$ slightly lowers the value of $H_0$ if we compare corresponding cases for the treatment of the redshift evolution.

    \item Independently on the likelihood, we can identify, similar to the case of only QSOs, a trend between the three cases with fixed correction for evolution and the case without correction. Indeed, the inclusion of the correction leads to higher uncertainties of the coefficients $a_i$ of the highest orders (mainly $a_4$ and $a_5$), due to the fact that also the additional uncertainties on the evolutionary parameters are taken into account. Moreover, since the correction for the redshift evolution affects more the high redshifts, the cases in which this correction is applied lead to best-fit values of the highest coefficients significantly discrepant from the ones obtained without accounting for the correction for the evolution. However, this discrepancy is partially alleviated by the large errors on the $a_i$ of the highest orders.
    From the point of view of this comparison between cases with and without correction, it is also visible that among the different applications of the correction for the redshift evolution, the one that assumes $\Omega_M=0.9$ is the one closer, in terms of best-fit values, to the case without evolution, as outlined in the case of QSOs alone. Once again, these two cases are the ones which lead to the lowest uncertainties on $a_3$, $a_4$, and $a_5$.
    
    \item Concerning the $a_i$ coefficients, we obtain best-fit values of $a_2$ compatible within 1 $\sigma$ among all evolutionary cases for each likelihood studied, while $a_3$, $a_4$, and $a_5$ show discrepancies among each other at different significance levels when comparing different treatments of the redshift evolution. Indeed, this discrepancy reaches a maximum of 5.7 $\sigma$ and 7 $\sigma$, for $a_4$ with $\cal L$$_{Gaussian}$ and $\cal L$$_{new}$, respectively, between the cases without correction and with correction and $\Omega_M=0.1$. Similarly, if we compare the cases without correction and with correction and $\Omega_M=0.3$, the maximum discrepancy is obtained for $a_4$ at a significance level of 4.1 $\sigma$ and 5.3 $\sigma$, with $\cal L$$_{Gaussian}$ and $\cal L$$_{new}$, respectively. For the cases without correction and with correction and $\Omega_M=0.9$, $a_4$ shows a discrepancy, which is of 3 $\sigma$ and 3.8 $\sigma$, with $\cal L$$_{Gaussian}$ and $\cal L$$_{new}$, respectively. 
    When comparing instead the best-fit values of the coefficients $a_i$ of the same order for the two different likelihoods and the same evolutionary corrections, we observe that $a_3$, $a_4$, and $a_5$ are compatible within 1 $\sigma$, while $a_2$ shows a discrepancy $\geq 4 \, \sigma$, also due to the lower uncertainties on $a_2$ compared to the ones of the other coefficients.

    \item Following Eqs. \eqref{coefflcdm_5}, we can also quantify the tension with the prediction of a flat $\Lambda$CDM model by comparing the best-fit $a_i$ values in Table \ref{tab:bestfit_all} with the values reported in Table \ref{tab:lcdm_all} for the corresponding cases. The only cases in which the obtained values are compatible within 1 $\sigma$ with the theoretical predictions are the following: $a_4$ in the case of correction for redshift evolution with $\Omega_M=0.9$ and $\cal L$$_{Gaussian}$ (at 0.9 $\sigma$ level), $a_4$ and $a_5$ in the case of correction for redshift evolution with $\Omega_M=0.9$ and $\cal L$$_{new}$ (at 0.8 and 0.1 $\sigma$ level respectively). All the other cases manifest a tension with the flat $\Lambda$CDM model $\geq 1 \, \sigma$, reaching a maximum discrepancy $> 30 \, \sigma$ on $a_2$ in the case of correction for redshift evolution with $\Omega_M=0.9$ for both the likelihoods.

    \item Based on the previous points, we can now compare our results to previous works in the literature that apply QSOs in cosmography together with other probes.
    In this regard, \citet{lusso2019} used a less updated sample of 1598 QSOs between $z=0.04$ and $z=5.1$, SNe Ia from the \textit{Pantheon sample} \citep{scolnic2018}, and 162 GRBs in the range $z=0.03 - 6.67$ standardized through the Amati relation. In \citet{lusso2019} work, the cosmographic model is a fourth-order non-orthogonal logarithmic polynomial and only the Gaussian likelihood is employed without any correction for the redshift evolution of luminosities of QSOs and GRBs. Beside these differences, they still found a strong overall discrepancy ($> 4 \, \sigma$) with a flat $\Lambda$CDM with $\Omega_M=0.3$. Moreover, in our study, thank to the orthogonalization of the cosmographic polynomial, we reduce the uncertainties on $a_2$, $a_3$, and $a_4$ compared to \citet{lusso2019}.\\
    The same cosmographic model (Eq. \eqref{Dlog}) as in our study, was instead already applied in \citet{2021A&A...649A..65B}, where \textit{Pantheon} SNe Ia were combined to the QSO sample used here but with only sources at $z>0.7$. In \citet{2021A&A...649A..65B} analysis, only the case without correction for evolution, with Gaussian likelihood, and with $H_0$ fixed to $70 \, \mathrm{km} \, \mathrm{s}^{-1} \, \mathrm{Mpc}^{-1}$ was considered. As in \citet{lusso2019}, the cosmographic fit of \citet{2021A&A...649A..65B} revealed a discrepancy $> 4 \, \sigma$ with a flat $\Lambda$CDM with $\Omega_M=0.3$. Comparing corresponding cases at $\Omega_M=0.3$, besides the differences in the data set and the analysis, we find compatibility within 1 $\sigma$ between our best-fit $a_i$ values and the ones reported in \citet{2021A&A...649A..65B}. This confirms the overall outcome of the cosmographic analysis, but we here stress that our work presents important points of novelty: the combination of the most updated samples of probes, the use of the QSO sample in the whole redshift range of observations which leverages also on low-redshift sources, the correction for redshift evolution and selection biases, and the employment of the best-fit likelihoods ($\cal L$$_{new}$). All these new points, combined together, proved to be crucial to better constrain the cosmographic model, thus providing a more precise description of the Universe, which is completely model-independent, and an improved estimate of the tension with the flat $\Lambda$CDM model.  
\end{itemize}

\begin{table*}
\caption{Best-fit values of the $a_i$ coefficients along with their 1 $\sigma$ uncertainty from QSOs alone for all cases investigated. $H_0$ is in units of $\, \mathrm{km} \, \mathrm{s}^{-1} \, \mathrm{Mpc}^{-1}$, but it is not constrained by QSOs alone.}
\begin{centering}
\begin{adjustbox}{width=\textwidth,center}
\begin{tabular}{ccccccccc}
\hline
\multicolumn{5}{c}{$\cal L$$_{Gaussian}$}&\multicolumn{4}{c}{$\cal L$$_{new}$}\tabularnewline
\hline
\hline
 & $a_2$ & $a_3$ & $a_4$ & $H_0$ & $a_2$ & $a_3$ & $a_4$ & $H_0$
\tabularnewline
\hline
No Evolution & $1.12 \pm 0.37$ & $-2.83 \pm 1.62$ & $10.53 \pm 6.31$ & $74.73 \pm 14.45$ & $1.19 \pm 0.38$ & $-3.18 \pm 1.56$ & $11.38 \pm 6.00$ & $74.94 \pm 15.50$ \tabularnewline
\hline
Fixed Evolution ($\Omega_M=0.3$) & $ 10.23\pm 2.82 $ & $ 9.56\pm 2.19$ & $29.70\pm 18.43$ & $ 75.24 \pm 14.51$ & $13.64 \pm 2.89$ & $10.12 \pm 2.04$ & $27.30 \pm 17.53$ & $75.97 \pm 14.45$ 
\tabularnewline
\hline
Fixed Evolution ($\Omega_M=0.1$) & $ 8.50\pm 3.38 $ & $19.08 \pm 3.27 $ & $52.57 \pm 27.82 $ &$75.61\pm 14.49$ & $13.86 \pm 3.52$ & $20.24 \pm 3.14$ & $48.56 \pm 27.09$ & $74.63 \pm 14.35$  \tabularnewline
\hline
Fixed Evolution ($\Omega_M=0.9$) & $11.57 \pm 1.94$ & $3.14 \pm 2.39$ & $21.59 \pm 18.00$ & $74.51 \pm 14.59$ & $12.61 \pm 2.00$ & $3.19 \pm 2.09$ & $17.53 \pm 13.91$ & $74.26 \pm 14.22$ \tabularnewline
\hline
\end{tabular}
\end{adjustbox}
\label{tab:bestfit_QSO}
\par\end{centering}
\end{table*}

\begin{table*}
\caption{Values of $a_i$ predicted in a flat $\Lambda$CDM model according to Eqs.  \eqref{coefflcdm_4} for only QSOs and all cases studied in this work.}
\begin{centering}
\begin{tabular}{cccccccc}
\hline
\multicolumn{1}{c}{}&\multicolumn{3}{c}{$\cal L$$_{Gaussian}$}&\multicolumn{3}{c}{$\cal L$$_{new}$}\tabularnewline
\hline
\hline
 & $a_2$ & $a_3$ & $a_4$ & $a_2$ & $a_3$ & $a_4$
\tabularnewline
\hline
Flat $\Lambda$CDM, No Evolution & 6.45 & 4.93 & 1.12 & 6.48 & 4.94 & 1.12  \tabularnewline
\hline
Flat $\Lambda$CDM, Fixed Evolution ($\Omega_M=0.3$) &  3.81& 4.40  & 1.12  & 5.03 & 4.40 &  1.12
\tabularnewline
\hline
Flat $\Lambda$CDM, Fixed Evolution ($\Omega_M=0.1$) &  2.51 & 8.21 & 4.57 & 5.13 & 8.19 & 4.57  \tabularnewline
\hline
Flat $\Lambda$CDM, Fixed Evolution ($\Omega_M=0.9$) & 4.46 & 2.43 & 1.08 & 4.51 & 2.43 & 1.08 \tabularnewline
\hline
\end{tabular}
\label{tab:lcdm_QSO}
\par\end{centering}
\end{table*}

\begin{table*}
\caption{Best-fit values of the $a_i$ coefficients along with their 1 $\sigma$ uncertainty from all probes together for all cases investigated. $H_0$ is in units of $\, \mathrm{km} \, \mathrm{s}^{-1} \, \mathrm{Mpc}^{-1}$.}
\begin{centering}
\begin{adjustbox}{width=\textwidth,center}
\begin{tabular}{ccccccccccc}
\hline
\multicolumn{6}{c}{$\cal L$$_{Gaussian}$}&\multicolumn{5}{c}{$\cal L$$_{new}$}\tabularnewline
\hline
\hline
 & $a_2$ & $a_3$ & $a_4$ & $a_5$ & $H_0$ & $a_2$ & $a_3$ & $a_4$ & $a_5$ & $H_0$
\tabularnewline
\hline
No Evolution & $3.90 \pm 0.05$ & $2.40 \pm 0.38$ & $-9.40 \pm 1.57$ & $9.69 \pm 5.83$ & $73.19 \pm 0.37$ & $3.56 \pm 0.04$ & $2.57 \pm 0.38$ & $-7.89 \pm 1.31$ & $3.88 \pm 5.60$ & $72.77 \pm 0.27$ \tabularnewline
\hline
Fixed Evolution ($\Omega_M=0.3$) & $3.98 \pm 0.05$ & $5.37 \pm 0.41$ & $5.77 \pm 2.13$ & $22.17 \pm 10.25$ & $73.54 \pm 0.38$ & $3.59 \pm 0.04$ & $4.84 \pm 0.36$& $8.63 \pm 1.89$ & $7.99 \pm 8.70$ & $73.11 \pm 0.28$
\tabularnewline
\hline
Fixed Evolution ($\Omega_M=0.1$) & $4.01 \pm 0.05$ & $5.56 \pm 0.45$ & $13.21 \pm 2.42$ & $35.76 \pm 11.93$ & $73.57 \pm 0.39$ & $3.63 \pm 0.04$ & $5.41 \pm 0.38$ & $16.85 \pm 2.24$ & $21.22 \pm 10.19$ & $73.16 \pm 0.28$ \tabularnewline
\hline
Fixed Evolution ($\Omega_M=0.9$) & $3.96 \pm 0.05$ & $4.37 \pm 0.40$ & $0.06 \pm 1.88$ & $13.59 \pm 8.66$ & $73.48 \pm 0.37$ & $3.62 \pm 0.04$ & $4.18 \pm 0.32$ & $3.00 \pm 1.63$ & $1.16 \pm 6.96$ & $73.06 \pm 0.27$ \tabularnewline
\hline
\end{tabular}
\end{adjustbox}
\label{tab:bestfit_all}
\par\end{centering}
\end{table*}

\begin{table*}
\caption{Values of $a_i$ predicted in a flat $\Lambda$CDM model according to Eqs.  \eqref{coefflcdm_5} for all probes together and all cases studied in this work.}
\begin{centering}
\begin{tabular}{cccccccccc}
\hline
\multicolumn{1}{c}{}&\multicolumn{4}{c}{$\cal L$$_{Gaussian}$}&\multicolumn{4}{c}{$\cal L$$_{new}$}\tabularnewline
\hline
\hline
 & $a_2$ & $a_3$ & $a_4$ & $a_5$ & $a_2$ & $a_3$ & $a_4$ & $a_5$
\tabularnewline
\hline
Flat $\Lambda$CDM, No Evolution & 3.76 & 3.67 & -0.85 & -1.82 & 3.65 & 3.68 & -0.91 & -1.82 \tabularnewline
\hline
Flat $\Lambda$CDM, Fixed Evolution ($\Omega_M=0.3$) & 3.76 & 3.68 & -0.55 & -1.82 & 3.68 & 3.69 & -0.52 & -1.82 
\tabularnewline
\hline
Flat $\Lambda$CDM, Fixed Evolution ($\Omega_M=0.1$) & 4.67 & 7.67 &5.40 & 0.94 & 4.56 & 7.41 & 5.39 & 0.94  \tabularnewline
\hline
Flat $\Lambda$CDM, Fixed Evolution ($\Omega_M=0.9$) & 2.29 & 2.15 & 1.66 & 0.61 & 2.27 & 2.17 & 1.67 & 0.61 \tabularnewline
\hline
\end{tabular}
\label{tab:lcdm_all}
\par\end{centering}
\end{table*}

\section{Conclusions}
\label{conclusions}

In this work, we have performed a completely model-independent test of the flat $\Lambda$CDM model leveraging the advantages of the robust cosmographic technique developed in \citet{2021A&A...649A..65B}, which is based on an orthogonal logarithmic polynomial of the luminosity distance. 
We have applied this procedure both to QSOs alone and to the combination of QSOs, GRBs, SNe Ia, and BAOs. In both cases we have also accounted for different treatments of the likelihoods \citep{Bargiacchi2023arXiv230307076B,Dainotti2023arXiv230510030D,snelikelihood2022_all}, and selection biases and redshift evolution of the luminosities of QSOs and GRBs \citep{Dainotti2013a,dainotti2015b,Dainotti2017A&A...600A..98D,Dainotti2021Galax...9...95D,Dainotti2022MNRAS.tmp.2639D,Bargiacchi2023arXiv230307076B,Dainotti2023arXiv230510030D,DainottiQSO,biasfreeQSO2022,Bargiacchi2023arXiv230307076B,Dainotti2023arXiv230510030D,DainottiGoldQSOApJ2023}. Indeed, we have distinguished four approaches for the correction for redshift evolution, one without correction and three different cases of correction when $\Omega_M$ is fixed to $0.1$, $0.3$ and $0.9$, and two approaches for the likelihoods, one with the traditional Gaussian and another one with the recently found best-fit likelihoods for each probe. 
Our comprehensive analysis leverages the cosmology independence of cosmography, the invaluable role of high-redshift sources, such as QSOs and GRBs, which are able to unveil tensions between the observed data and the predictions of cosmological models, and the power of low-redshift probes, such as SNe Ia and BAOs, which lead to an increased precision in the determination of the cosmological parameters.

Our study shows that QSOs alone still needs to be improved, at least for the total sample of 2421 sources here presented, compared to the combination of all probes, leading to large uncertainties on the cosmographic free parameters and a $H_0$ not constrained. Nonetheless, they provide a coverage of the high redshift range that brings crucial information on the evolution of the Universe and thus on the discrepancies with cosmological models pointing toward a tension with the flat $\Lambda$CDM model.
Once QSOs are combined with GRBs, SNe Ia, and BAOs, all parameters are better constrained and $H_0$, driven by SNe Ia and BAOs, is in all cases compatible with the most recent value reported in \citet{2022ApJ...934L...7R}.
Both when dealing with only QSOs and with all probes together, the inclusion of the correction for redshift evolution significantly changes the results, while the application of $\cal L$$_{new}$ strongly reduces the uncertainties on parameters compared to $\cal L$$_{Gaussian}$, leading to a maximum reduction on $H_0$ of $\sim 28 \%$. This marks the importance of accounting for the redshift evolution and the best-fit likelihood to obtain the actual best-fits with better constraints. 
These results of better constraining the cosmographic parameters is indeed alligned with the reduction of the scatter we have on the cosmological parameters when we use the $\cal L$$_{new}$. 
These results also stress the importance of these new likelihoods both in the cosmology and cosmographic domain.
Finally, in all the cases investigated, we have obtained an overall discrepancy between the data and the flat $\Lambda$CDM model, which in some cases reaches a statistical significance, in agreement with previous cosmographic studies \citep{lusso2019,2021A&A...649A..65B}.
On the other hand, the tension achieved in this work reaches a statistical level much higher than the ones obtained in cosmological studies which use the combination of QSOs, GRBs, SNe Ia, and BAOs, such as \citet{Bargiacchi2023arXiv230307076B} and \citet{Dainotti2023arXiv230510030D}. Specifically, \citet{Bargiacchi2023arXiv230307076B} have investigated the flat $\Lambda$CDM model, while \citet{Dainotti2023arXiv230510030D} have extended the analysis to the non-flat $\Lambda$CDM and flat $w$CDM model. Thus, the fact that in this study the tension proves to be increased, compared to those works, means that the cosmographic model points toward a cosmological model different (especially phantom model, \citealt{biasfreeQSO2022}) to the ones already investigated. 
Another important point to be taken into account is that in this treatment, differently from the cosmological approach, we cannot apply the evolutionary function which changes at the change of $\Omega_M$ \citep{biasfreeQSO2022,Bargiacchi2023arXiv230307076B,Dainotti2023arXiv230510030D} and prevents the 
values of $\Omega_M$ to be closer to 1. 
Thus, if a cure to this effect could indeed be used, the values of $\Omega_M$ would be shifted at smaller values and we envision the discrepancy could be smaller.
In conclusion, we have demonstrated the power of the cosmographic approach embedded with high-redshift probes, once they are carefully corrected for biases and redshift evolution and the proper likelihoods are employed. 

In this scenario, without any specific cosmological assumption, a tension has emerged between the observations and the theoretical prediction of the flat $\Lambda$CDM model, making the hunt for the most suitable description of the Universe even more puzzling along with the quest of the reasons for these discrepancies.

\section*{Acknowledgements}

GB acknowledges Scuola Superiore Meridionale, for supporting her visit at NAOJ, Division of Science. GB thanks the Division of Science for being hosted on campus.

\section*{Data Availability}

The data underlying this article will be shared upon a reasonable request to the corresponding author.



\bibliographystyle{mnras}
\bibliography{bibliografia} 

\begin{thebibliography}{}
\makeatletter
\relax
\def\mn@urlcharsother{\let\do\@makeother \do\$\do\&\do\#\do\^\do\_\do\%\do\~}
\def\mn@doi{\begingroup\mn@urlcharsother \@ifnextchar [ {\mn@doi@}
  {\mn@doi@[]}}
\def\mn@doi@[#1]#2{\def\@tempa{#1}\ifx\@tempa\@empty \href
  {http://dx.doi.org/#2} {doi:#2}\else \href {http://dx.doi.org/#2} {#1}\fi
  \endgroup}
\def\mn@eprint#1#2{\mn@eprint@#1:#2::\@nil}
\def\mn@eprint@arXiv#1{\href {http://arxiv.org/abs/#1} {{\tt arXiv:#1}}}
\def\mn@eprint@dblp#1{\href {http://dblp.uni-trier.de/rec/bibtex/#1.xml}
  {dblp:#1}}
\def\mn@eprint@#1:#2:#3:#4\@nil{\def\@tempa {#1}\def\@tempb {#2}\def\@tempc
  {#3}\ifx \@tempc \@empty \let \@tempc \@tempb \let \@tempb \@tempa \fi \ifx
  \@tempb \@empty \def\@tempb {arXiv}\fi \@ifundefined
  {mn@eprint@\@tempb}{\@tempb:\@tempc}{\expandafter \expandafter \csname
  mn@eprint@\@tempb\endcsname \expandafter{\@tempc}}}

\bibitem[\protect\citeauthoryear{{Aubourg} et~al.,}{{Aubourg}
  et~al.}{2015}]{2015PhRvD..92l3516A}
{Aubourg} {\'E}.,  et~al., 2015, \mn@doi [\prd] {10.1103/PhysRevD.92.123516},
  \href {https://ui.adsabs.harvard.edu/abs/2015PhRvD..92l3516A} {92, 123516}

\bibitem[\protect\citeauthoryear{{Aviles}, {Bravetti}, {Capozziello}  \&
  {Luongo}}{{Aviles} et~al.}{2014}]{2014PhRvD..90d3531A}
{Aviles} A.,  {Bravetti} A.,  {Capozziello} S.,   {Luongo} O.,  2014, \mn@doi
  [\prd] {10.1103/PhysRevD.90.043531}, \href
  {https://ui.adsabs.harvard.edu/abs/2014PhRvD..90d3531A} {90, 043531}

\bibitem[\protect\citeauthoryear{{Avni} \& {Tananbaum}}{{Avni} \&
  {Tananbaum}}{1986}]{1986ApJ...305...83A}
{Avni} Y.,  {Tananbaum} H.,  1986, \mn@doi [\apj] {10.1086/164230}, \href
  {https://ui.adsabs.harvard.edu/abs/1986ApJ...305...83A} {305, 83}

\bibitem[\protect\citeauthoryear{{Bargiacchi}, {Risaliti}, {Benetti},
  {Capozziello}, {Lusso}, {Saccardi}  \& {Signorini}}{{Bargiacchi}
  et~al.}{2021}]{2021A&A...649A..65B}
{Bargiacchi} G.,  {Risaliti} G.,  {Benetti} M.,  {Capozziello} S.,  {Lusso} E.,
   {Saccardi} A.,   {Signorini} M.,  2021, \mn@doi [\aap]
  {10.1051/0004-6361/202140386}, \href
  {https://ui.adsabs.harvard.edu/abs/2021A&A...649A..65B} {649, A65}

\bibitem[\protect\citeauthoryear{{Bargiacchi}, {Benetti}, {Capozziello},
  {Lusso}, {Risaliti}  \& {Signorini}}{{Bargiacchi}
  et~al.}{2022}]{2022MNRAS.515.1795B}
{Bargiacchi} G.,  {Benetti} M.,  {Capozziello} S.,  {Lusso} E.,  {Risaliti} G.,
    {Signorini} M.,  2022, \mn@doi [\mnras] {10.1093/mnras/stac1941}, \href
  {https://ui.adsabs.harvard.edu/abs/2022MNRAS.515.1795B} {515, 1795}

\bibitem[\protect\citeauthoryear{{Bargiacchi}, {Dainotti}, {Nagataki}  \&
  {Capozziello}}{{Bargiacchi} et~al.}{2023}]{Bargiacchi2023arXiv230307076B}
{Bargiacchi} G.,  {Dainotti} M.~G.,  {Nagataki} S.,   {Capozziello} S.,  2023,
  \mn@doi [\mnras] {10.1093/mnras/stad763}, \href
  {https://ui.adsabs.harvard.edu/abs/2023MNRAS.tmp..761B} {}

\bibitem[\protect\citeauthoryear{Benetti \& Capozziello}{Benetti \&
  Capozziello}{2019}]{Benetti}
Benetti M.,  Capozziello S.,  2019, \mn@doi [JCAP]
  {10.1088/1475-7516/2019/12/008}, 12, 008

\bibitem[\protect\citeauthoryear{{Bernardini}, {Margutti}, {Zaninoni}  \&
  {Chincarini}}{{Bernardini} et~al.}{2012}]{2012MNRAS.425.1199B}
{Bernardini} M.~G.,  {Margutti} R.,  {Zaninoni} E.,   {Chincarini} G.,  2012,
  \mn@doi [\mnras] {10.1111/j.1365-2966.2012.21487.x}, \href
  {https://ui.adsabs.harvard.edu/abs/2012MNRAS.425.1199B} {425, 1199}

\bibitem[\protect\citeauthoryear{{Beutler} et~al.,}{{Beutler}
  et~al.}{2011}]{2011MNRAS.416.3017B}
{Beutler} F.,  et~al., 2011, \mn@doi [\mnras]
  {10.1111/j.1365-2966.2011.19250.x}, \href
  {https://ui.adsabs.harvard.edu/abs/2011MNRAS.416.3017B} {416, 3017}

\bibitem[\protect\citeauthoryear{{Bisogni}, {Lusso}, {Civano}, {Nardini},
  {Risaliti}, {Elvis}  \& {Fabbiano}}{{Bisogni}
  et~al.}{2021}]{2021A&A...655A.109B}
{Bisogni} S.,  {Lusso} E.,  {Civano} F.,  {Nardini} E.,  {Risaliti} G.,
  {Elvis} M.,   {Fabbiano} G.,  2021, \mn@doi [\aap]
  {10.1051/0004-6361/202140852}, \href
  {https://ui.adsabs.harvard.edu/abs/2021A&A...655A.109B} {655, A109}

\bibitem[\protect\citeauthoryear{{Bloom}, {Frail}  \& {Sari}}{{Bloom}
  et~al.}{2001}]{2001AJ....121.2879B}
{Bloom} J.~S.,  {Frail} D.~A.,   {Sari} R.,  2001, \mn@doi [\aj]
  {10.1086/321093}, \href
  {https://ui.adsabs.harvard.edu/abs/2001AJ....121.2879B} {121, 2879}

\bibitem[\protect\citeauthoryear{{Brout} et~al.,}{{Brout}
  et~al.}{2022}]{2022ApJ...938..110B}
{Brout} D.,  et~al., 2022, \mn@doi [\apj] {10.3847/1538-4357/ac8e04}, \href
  {https://ui.adsabs.harvard.edu/abs/2022ApJ...938..110B} {938, 110}

\bibitem[\protect\citeauthoryear{{Capozziello}, {D'Agostino}  \&
  {Luongo}}{{Capozziello} et~al.}{2018}]{2018MNRAS.476.3924C}
{Capozziello} S.,  {D'Agostino} R.,   {Luongo} O.,  2018, \mn@doi [\mnras]
  {10.1093/mnras/sty422}, \href
  {https://ui.adsabs.harvard.edu/abs/2018MNRAS.476.3924C} {476, 3924}

\bibitem[\protect\citeauthoryear{{Capozziello}, {D'Agostino}  \&
  {Luongo}}{{Capozziello} et~al.}{2019}]{2019IJMPD..2830016C}
{Capozziello} S.,  {D'Agostino} R.,   {Luongo} O.,  2019, \mn@doi
  [International Journal of Modern Physics D] {10.1142/S0218271819300167},
  \href {https://ui.adsabs.harvard.edu/abs/2019IJMPD..2830016C} {28, 1930016}

\bibitem[\protect\citeauthoryear{{Capozziello}, {D'Agostino}  \&
  {Luongo}}{{Capozziello} et~al.}{2020}]{2020MNRAS.494.2576C}
{Capozziello} S.,  {D'Agostino} R.,   {Luongo} O.,  2020, \mn@doi [\mnras]
  {10.1093/mnras/staa871}, \href
  {https://ui.adsabs.harvard.edu/abs/2020MNRAS.494.2576C} {494, 2576}

\bibitem[\protect\citeauthoryear{{Cardone}, {Capozziello}  \&
  {Dainotti}}{{Cardone} et~al.}{2009}]{cardone09}
{Cardone} V.~F.,  {Capozziello} S.,   {Dainotti} M.~G.,  2009, \mn@doi [\mnras]
  {10.1111/j.1365-2966.2009.15456.x}, \href
  {http://adsabs.harvard.edu/abs/2009MNRAS.400..775C} {400, 775}

\bibitem[\protect\citeauthoryear{{Cardone}, {Dainotti}, {Capozziello}  \&
  {Willingale}}{{Cardone} et~al.}{2010}]{cardone10}
{Cardone} V.~F.,  {Dainotti} M.~G.,  {Capozziello} S.,   {Willingale} R.,
  2010, \mn@doi [\mnras] {10.1111/j.1365-2966.2010.17197.x}, \href
  {http://adsabs.harvard.edu/abs/2010MNRAS.408.1181C} {408, 1181}

\bibitem[\protect\citeauthoryear{{Carroll}}{{Carroll}}{2001}]{2001LRR.....4....1C}
{Carroll} S.~M.,  2001, \mn@doi [Living Reviews in Relativity]
  {10.12942/lrr-2001-1}, \href
  {https://ui.adsabs.harvard.edu/abs/2001LRR.....4....1C} {4, 1}

\bibitem[\protect\citeauthoryear{{Colg{\'a}in}, {Sheikh-Jabbari}, {Solomon},
  {Bargiacchi}, {Capozziello}, {Dainotti}  \& {Stojkovic}}{{Colg{\'a}in}
  et~al.}{2022a}]{2022arXiv220310558C}
{Colg{\'a}in} E.~{\'O}.,  {Sheikh-Jabbari} M.~M.,  {Solomon} R.,  {Bargiacchi}
  G.,  {Capozziello} S.,  {Dainotti} M.~G.,   {Stojkovic} D.,  2022a, \mn@doi
  [arXiv e-prints] {10.48550/arXiv.2203.10558}, \href
  {https://ui.adsabs.harvard.edu/abs/2022arXiv220310558C} {p. arXiv:2203.10558}

\bibitem[\protect\citeauthoryear{{Colg{\'a}in}, {Sheikh-Jabbari}, {Solomon},
  {Dainotti}  \& {Stojkovic}}{{Colg{\'a}in}
  et~al.}{2022b}]{Colgain2022arXiv220611447C}
{Colg{\'a}in} E.~{\'O}.,  {Sheikh-Jabbari} M.~M.,  {Solomon} R.,  {Dainotti}
  M.~G.,   {Stojkovic} D.,  2022b, arXiv e-prints, \href
  {https://ui.adsabs.harvard.edu/abs/2022arXiv220611447C} {p. arXiv:2206.11447}

\bibitem[\protect\citeauthoryear{{Cucchiara} et~al.,}{{Cucchiara}
  et~al.}{2011}]{2011ApJ...736....7C}
{Cucchiara} A.,  et~al., 2011, \mn@doi [\apj] {10.1088/0004-637X/736/1/7},
  \href {https://ui.adsabs.harvard.edu/abs/2011ApJ...736....7C} {736, 7}

\bibitem[\protect\citeauthoryear{{Cuceu}, {Farr}, {Lemos}  \&
  {Font-Ribera}}{{Cuceu} et~al.}{2019}]{2019JCAP...10..044C}
{Cuceu} A.,  {Farr} J.,  {Lemos} P.,   {Font-Ribera} A.,  2019, \mn@doi [\jcap]
  {10.1088/1475-7516/2019/10/044}, \href
  {https://ui.adsabs.harvard.edu/abs/2019JCAP...10..044C} {2019, 044}

\bibitem[\protect\citeauthoryear{{Dainotti} \& {Amati}}{{Dainotti} \&
  {Amati}}{2018}]{Dainotti2018PASP..130e1001D}
{Dainotti} M.~G.,  {Amati} L.,  2018, \mn@doi [\pasp]
  {10.1088/1538-3873/aaa8d7}, \href
  {https://ui.adsabs.harvard.edu/abs/2018PASP..130e1001D} {130, 051001}

\bibitem[\protect\citeauthoryear{{Dainotti} \& {Del Vecchio}}{{Dainotti} \&
  {Del Vecchio}}{2017}]{Dainotti2017NewAR..77...23D}
{Dainotti} M.~G.,  {Del Vecchio} R.,  2017, \mn@doi [\nar]
  {10.1016/j.newar.2017.04.001}, \href
  {https://ui.adsabs.harvard.edu/abs/2017NewAR..77...23D} {77, 23}

\bibitem[\protect\citeauthoryear{{Dainotti}, {Cardone}  \&
  {Capozziello}}{{Dainotti} et~al.}{2008}]{Dainotti2008}
{Dainotti} M.~G.,  {Cardone} V.~F.,   {Capozziello} S.,  2008, \mn@doi [\mnras]
  {10.1111/j.1745-3933.2008.00560.x}, \href
  {http://adsabs.harvard.edu/abs/2008MNRAS.391L..79D} {391, L79}

\bibitem[\protect\citeauthoryear{{Dainotti}, {Willingale}, {Capozziello},
  {Fabrizio Cardone}  \& {Ostrowski}}{{Dainotti}
  et~al.}{2010}]{Dainotti2010ApJ...722L.215D}
{Dainotti} M.~G.,  {Willingale} R.,  {Capozziello} S.,  {Fabrizio Cardone} V.,
   {Ostrowski} M.,  2010, \mn@doi [\apjl] {10.1088/2041-8205/722/2/L215}, \href
  {https://ui.adsabs.harvard.edu/abs/2010ApJ...722L.215D} {722, L215}

\bibitem[\protect\citeauthoryear{{Dainotti}, {Fabrizio Cardone}, {Capozziello},
  {Ostrowski}  \& {Willingale}}{{Dainotti} et~al.}{2011}]{dainotti11a}
{Dainotti} M.~G.,  {Fabrizio Cardone} V.,  {Capozziello} S.,  {Ostrowski} M.,
  {Willingale} R.,  2011, \mn@doi [\apj] {10.1088/0004-637X/730/2/135}, \href
  {http://adsabs.harvard.edu/abs/2011ApJ...730..135D} {730, 135}

\bibitem[\protect\citeauthoryear{{Dainotti}, {Cardone}, {Piedipalumbo}  \&
  {Capozziello}}{{Dainotti} et~al.}{2013a}]{Dainotti2013a}
{Dainotti} M.~G.,  {Cardone} V.~F.,  {Piedipalumbo} E.,   {Capozziello} S.,
  2013a, \mn@doi [\mnras] {10.1093/mnras/stt1516}, \href
  {http://adsabs.harvard.edu/abs/2013MNRAS.436...82D} {436, 82}

\bibitem[\protect\citeauthoryear{{Dainotti}, {Petrosian}, {Singal}  \&
  {Ostrowski}}{{Dainotti} et~al.}{2013b}]{Dainotti2013b}
{Dainotti} M.~G.,  {Petrosian} V.,  {Singal} J.,   {Ostrowski} M.,  2013b,
  \mn@doi [\apj] {10.1088/0004-637X/774/2/157}, \href
  {http://adsabs.harvard.edu/abs/2013ApJ...774..157D} {774, 157}

\bibitem[\protect\citeauthoryear{{Dainotti}, {Petrosian}, {Willingale},
  {O'Brien}, {Ostrowski}  \& {Nagataki}}{{Dainotti}
  et~al.}{2015}]{dainotti2015b}
{Dainotti} M.,  {Petrosian} V.,  {Willingale} R.,  {O'Brien} P.,  {Ostrowski}
  M.,   {Nagataki} S.,  2015, \mn@doi [\mnras] {10.1093/mnras/stv1229}, \href
  {https://ui.adsabs.harvard.edu/abs/2015MNRAS.451.3898D} {451, 3898}

\bibitem[\protect\citeauthoryear{{Dainotti}, {Postnikov}, {Hernandez}  \&
  {Ostrowski}}{{Dainotti} et~al.}{2016}]{Dainotti2016ApJ...825L..20D}
{Dainotti} M.~G.,  {Postnikov} S.,  {Hernandez} X.,   {Ostrowski} M.,  2016,
  \mn@doi [\apjl] {10.3847/2041-8205/825/2/L20}, \href
  {https://ui.adsabs.harvard.edu/abs/2016ApJ...825L..20D} {825, L20}

\bibitem[\protect\citeauthoryear{{Dainotti}, {Nagataki}, {Maeda}, {Postnikov}
  \& {Pian}}{{Dainotti} et~al.}{2017a}]{Dainotti2017A&A...600A..98D}
{Dainotti} M.~G.,  {Nagataki} S.,  {Maeda} K.,  {Postnikov} S.,   {Pian} E.,
  2017a, \mn@doi [\aap] {10.1051/0004-6361/201628384}, \href
  {https://ui.adsabs.harvard.edu/abs/2017A&A...600A..98D} {600, A98}

\bibitem[\protect\citeauthoryear{{Dainotti}, {Hernandez}, {Postnikov},
  {Nagataki}, {O'brien}, {Willingale}  \& {Striegel}}{{Dainotti}
  et~al.}{2017b}]{Dainotti2017ApJ...848...88D}
{Dainotti} M.~G.,  {Hernandez} X.,  {Postnikov} S.,  {Nagataki} S.,  {O'brien}
  P.,  {Willingale} R.,   {Striegel} S.,  2017b, \mn@doi [\apj]
  {10.3847/1538-4357/aa8a6b}, \href
  {https://ui.adsabs.harvard.edu/abs/2017ApJ...848...88D} {848, 88}

\bibitem[\protect\citeauthoryear{{Dainotti}, {Del Vecchio}  \&
  {Tarnopolski}}{{Dainotti} et~al.}{2018}]{Dainotti2018AdAst2018E...1D}
{Dainotti} M.~G.,  {Del Vecchio} R.,   {Tarnopolski} M.,  2018, \mn@doi
  [Advances in Astronomy] {10.1155/2018/4969503}, \href
  {https://ui.adsabs.harvard.edu/abs/2018AdAst2018E...1D} {2018, 4969503}

\bibitem[\protect\citeauthoryear{{Dainotti}, {Lenart}, {Sarracino}, {Nagataki},
  {Capozziello}  \& {Fraija}}{{Dainotti} et~al.}{2020a}]{Dainotti2020a}
{Dainotti} M.,  {Lenart} A.,  {Sarracino} G.,  {Nagataki} S.,  {Capozziello}
  S.,   {Fraija} N.,  2020a, \mn@doi [\apj] {doi:10.3847/1538-4357/abbe8a},
  \href {https://ui.adsabs.harvard.edu/abs/2020arXiv201002092D/abstract} {904,
  19}

\bibitem[\protect\citeauthoryear{{Dainotti}, {Lenart}, {Sarracino}, {Nagataki},
  {Capozziello}  \& {Fraija}}{{Dainotti}
  et~al.}{2020b}]{Dainotti2020ApJ...904...97D}
{Dainotti} M.~G.,  {Lenart} A.~{\L}.,  {Sarracino} G.,  {Nagataki} S.,
  {Capozziello} S.,   {Fraija} N.,  2020b, \mn@doi [\apj]
  {10.3847/1538-4357/abbe8a}, \href
  {https://ui.adsabs.harvard.edu/abs/2020ApJ...904...97D} {904, 97}

\bibitem[\protect\citeauthoryear{{Dainotti} et~al.,}{{Dainotti}
  et~al.}{2020c}]{Dainotti2020b}
{Dainotti} M.~G.,  et~al., 2020c, \mn@doi [\apjl] {10.3847/2041-8213/abcda9},
  \href {https://ui.adsabs.harvard.edu/abs/2020ApJ...905L..26D} {905, L26}

\bibitem[\protect\citeauthoryear{{Dainotti}, {Levine}, {Fraija}  \&
  {Chandra}}{{Dainotti} et~al.}{2021a}]{Dainotti2021Galax...9...95D}
{Dainotti} M.,  {Levine} D.,  {Fraija} N.,   {Chandra} P.,  2021a, \mn@doi
  [Galaxies] {10.3390/galaxies9040095}, \href
  {https://ui.adsabs.harvard.edu/abs/2021Galax...9...95D} {9, 95}

\bibitem[\protect\citeauthoryear{{Dainotti} et~al.,}{{Dainotti}
  et~al.}{2021b}]{Dainotti2021ApJS..255...13D}
{Dainotti} M.~G.,  et~al., 2021b, \mn@doi [\apjs] {10.3847/1538-4365/abfe17},
  \href {https://ui.adsabs.harvard.edu/abs/2021ApJS..255...13D} {255, 13}

\bibitem[\protect\citeauthoryear{{Dainotti}, {De Simone}, {Schiavone},
  {Montani}, {Rinaldi}  \& {Lambiase}}{{Dainotti}
  et~al.}{2021c}]{Dainotti2021ApJ...912..150D}
{Dainotti} M.~G.,  {De Simone} B.,  {Schiavone} T.,  {Montani} G.,  {Rinaldi}
  E.,   {Lambiase} G.,  2021c, \mn@doi [\apj] {10.3847/1538-4357/abeb73}, \href
  {https://ui.adsabs.harvard.edu/abs/2021ApJ...912..150D} {912, 150}

\bibitem[\protect\citeauthoryear{{Dainotti}, {Petrosian}  \&
  {Bowden}}{{Dainotti} et~al.}{2021d}]{Dainotti2021ApJ...914L..40D}
{Dainotti} M.~G.,  {Petrosian} V.,   {Bowden} L.,  2021d, \mn@doi [\apjl]
  {10.3847/2041-8213/abf5e4}, \href
  {https://ui.adsabs.harvard.edu/abs/2021ApJ...914L..40D} {914, L40}

\bibitem[\protect\citeauthoryear{{Dainotti}, {Sarracino}  \&
  {Capozziello}}{{Dainotti} et~al.}{2022b}]{Dainotti2022PASJ}
{Dainotti} M.~G.,  {Sarracino} G.,   {Capozziello} S.,  2022b, \mn@doi [\pasj]
  {10.1093/pasj/psac057}, \href
  {https://ui.adsabs.harvard.edu/abs/2022PASJ..tmp...83D} {}

\bibitem[\protect\citeauthoryear{{Dainotti}, {Lenart}, {Chraya}, {Sarracino},
  {Nagataki}, {Fraija}, {Capozziello}  \& {Bogdan}}{{Dainotti}
  et~al.}{2022a}]{Dainotti2022MNRAS.tmp.2639D}
{Dainotti} M.~G.,  {Lenart} A.~L.,  {Chraya} A.,  {Sarracino} G.,  {Nagataki}
  S.,  {Fraija} N.,  {Capozziello} S.,   {Bogdan} M.,  2022a, \mn@doi [\mnras]
  {10.1093/mnras/stac2752}, \href
  {https://ui.adsabs.harvard.edu/abs/2022MNRAS.tmp.2639D} {}

\bibitem[\protect\citeauthoryear{{Dainotti}, {De Simone}, {Schiavone},
  {Montani}, {Rinaldi}, {Lambiase}, {Bogdan}  \& {Ugale}}{{Dainotti}
  et~al.}{2022c}]{2022Galax..10...24D}
{Dainotti} M.~G.,  {De Simone} B.~D.,  {Schiavone} T.,  {Montani} G.,
  {Rinaldi} E.,  {Lambiase} G.,  {Bogdan} M.,   {Ugale} S.,  2022c, \mn@doi
  [Galaxies] {10.3390/galaxies10010024}, \href
  {https://ui.adsabs.harvard.edu/abs/2022Galax..10...24D} {10, 24}

\bibitem[\protect\citeauthoryear{{Dainotti}, {De Simone}, {Schiavone},
  {Montani}, {Rinaldi}, {Lambiase}, {Bogdan}  \& {Ugale}}{{Dainotti}
  et~al.}{2022d}]{Dainotti2022Galax..10...24D}
{Dainotti} M.~G.,  {De Simone} B.,  {Schiavone} T.,  {Montani} G.,  {Rinaldi}
  E.,  {Lambiase} G.,  {Bogdan} M.,   {Ugale} S.,  2022d, \mn@doi [Galaxies]
  {doi:10.3390/galaxies10010024}, 10, 24

\bibitem[\protect\citeauthoryear{{Dainotti}, {Sarracino}  \&
  {Capozziello}}{{Dainotti} et~al.}{2022e}]{Dainotti2022PASJ...74.1095D}
{Dainotti} M.~G.,  {Sarracino} G.,   {Capozziello} S.,  2022e, \mn@doi [\pasj]
  {10.1093/pasj/psac057}, \href
  {https://ui.adsabs.harvard.edu/abs/2022PASJ...74.1095D} {74, 1095}

\bibitem[\protect\citeauthoryear{{Dainotti} et~al.,}{{Dainotti}
  et~al.}{2022f}]{Dainotti2022ApJS..261...25D}
{Dainotti} M.~G.,  et~al., 2022f, \mn@doi [\apjs] {10.3847/1538-4365/ac7c64},
  \href {https://ui.adsabs.harvard.edu/abs/2022ApJS..261...25D} {261, 25}

\bibitem[\protect\citeauthoryear{{Dainotti}, {Nielson}, {Sarracino}, {Rinaldi},
  {Nagataki}, {Capozziello}, {Gnedin}  \& {Bargiacchi}}{{Dainotti}
  et~al.}{2022g}]{Dainotti2022MNRAS.514.1828D}
{Dainotti} M.~G.,  {Nielson} V.,  {Sarracino} G.,  {Rinaldi} E.,  {Nagataki}
  S.,  {Capozziello} S.,  {Gnedin} O.~Y.,   {Bargiacchi} G.,  2022g, \mn@doi
  [\mnras] {10.1093/mnras/stac1141}, \href
  {https://ui.adsabs.harvard.edu/abs/2022MNRAS.514.1828D} {514, 1828}

\bibitem[\protect\citeauthoryear{{Dainotti}, {Bargiacchi}, {Lenart},
  {Capozziello}, {{\'O} Colg{\'a}in}, {Solomon}, {Stojkovic}  \&
  {Sheikh-Jabbari}}{{Dainotti} et~al.}{2022h}]{DainottiQSO}
{Dainotti} M.~G.,  {Bargiacchi} G.,  {Lenart} A.~{\L}.,  {Capozziello} S.,
  {{\'O} Colg{\'a}in} E.,  {Solomon} R.,  {Stojkovic} D.,   {Sheikh-Jabbari}
  M.~M.,  2022h, \mn@doi [\apj] {10.3847/1538-4357/ac6593}, \href
  {https://ui.adsabs.harvard.edu/abs/2022ApJ...931..106D} {931, 106}

\bibitem[\protect\citeauthoryear{{Dainotti}, {Levine}, {Fraija}, {Warren}  \&
  {Sourav}}{{Dainotti}
  et~al.}{2022i}]{Dainotticlosureoptical2022ApJ...940..169D}
{Dainotti} M.~G.,  {Levine} D.,  {Fraija} N.,  {Warren} D.,   {Sourav} S.,
  2022i, \mn@doi [\apj] {10.3847/1538-4357/ac9b11}, \href
  {https://ui.adsabs.harvard.edu/abs/2022ApJ...940..169D} {940, 169}

\bibitem[\protect\citeauthoryear{{Dainotti}, {Bargiacchi}, {Bogdan},
  {Capozziello}  \& {Nagataki}}{{Dainotti}
  et~al.}{2023b}]{snelikelihood2022_all}
{Dainotti} M.~G.,  {Bargiacchi} G.,  {Bogdan} M.,  {Capozziello} S.,
  {Nagataki} S.,  2023b, JCAP submitted

\bibitem[\protect\citeauthoryear{Dainotti, Bargiacchi, Lenart, Nagataki  \&
  Capozziello}{Dainotti et~al.}{2023a}]{DainottiGoldQSOApJ2023}
Dainotti M.~G.,  Bargiacchi G.,  Lenart A.,  Nagataki S.,   Capozziello S.,
  2023a, \apj

\bibitem[\protect\citeauthoryear{{Dainotti}, {Bargiacchi}, {Bogdan}, {{\L}ukasz
  Lenart}, {Iwasaki}, {Capozziello}, {Zhang}  \& {Fraija}}{{Dainotti}
  et~al.}{2023c}]{Dainotti2023arXiv230510030D}
{Dainotti} M.~G.,  {Bargiacchi} G.,  {Bogdan} M.,  {{\L}ukasz Lenart} A.,
  {Iwasaki} K.,  {Capozziello} S.,  {Zhang} B.,   {Fraija} N.,  2023c, \mn@doi
  [arXiv e-prints] {10.48550/arXiv.2305.10030}, \href
  {https://ui.adsabs.harvard.edu/abs/2023arXiv230510030D} {p. arXiv:2305.10030}

\bibitem[\protect\citeauthoryear{{Dall'Osso}, {Stratta}, {Guetta}, {Covino},
  {De Cesare}  \& {Stella}}{{Dall'Osso} et~al.}{2011}]{2011A&A...526A.121D}
{Dall'Osso} S.,  {Stratta} G.,  {Guetta} D.,  {Covino} S.,  {De Cesare} G.,
  {Stella} L.,  2011, \mn@doi [\aap] {10.1051/0004-6361/201014168}, \href
  {https://ui.adsabs.harvard.edu/abs/2011A&A...526A.121D} {526, A121}

\bibitem[\protect\citeauthoryear{{Demianski}, {Piedipalumbo}, {Sawant}  \&
  {Amati}}{{Demianski} et~al.}{2017}]{2017A&A...598A.113D}
{Demianski} M.,  {Piedipalumbo} E.,  {Sawant} D.,   {Amati} L.,  2017, \mn@doi
  [\aap] {10.1051/0004-6361/201628911}, \href
  {https://ui.adsabs.harvard.edu/abs/2017A&A...598A.113D} {598, A113}

\bibitem[\protect\citeauthoryear{Di~Valentino, Melchiorri  \&
  Silk}{Di~Valentino et~al.}{2021}]{DiValentino:2020hov}
Di~Valentino E.,  Melchiorri A.,   Silk J.,  2021, \mn@doi [Astrophys. J.
  Lett.] {10.3847/2041-8213/abe1c4}, 908, L9

\bibitem[\protect\citeauthoryear{{Efron} \& {Petrosian}}{{Efron} \&
  {Petrosian}}{1992}]{1992ApJ...399..345E}
{Efron} B.,  {Petrosian} V.,  1992, \mn@doi [\apj] {10.1086/171931}, \href
  {https://ui.adsabs.harvard.edu/abs/1992ApJ...399..345E} {399, 345}

\bibitem[\protect\citeauthoryear{Efstathiou \& Gratton}{Efstathiou \&
  Gratton}{2020}]{Efstathiou:2020wem}
Efstathiou G.,  Gratton S.,  2020, \mn@doi [Mon. Not. Roy. Astron. Soc.]
  {10.1093/mnrasl/slaa093}, 496, L91

\bibitem[\protect\citeauthoryear{{Eisenstein} et~al.,}{{Eisenstein}
  et~al.}{2005}]{2005ApJ...633..560E}
{Eisenstein} D.~J.,  et~al., 2005, \mn@doi [\apj] {10.1086/466512}, \href
  {https://ui.adsabs.harvard.edu/abs/2005ApJ...633..560E} {633, 560}

\bibitem[\protect\citeauthoryear{{Escamilla-Rivera} \&
  {Capozziello}}{{Escamilla-Rivera} \&
  {Capozziello}}{2019}]{2019IJMPD..2850154E}
{Escamilla-Rivera} C.,  {Capozziello} S.,  2019, \mn@doi [International Journal
  of Modern Physics D] {10.1142/S0218271819501542}, \href
  {https://ui.adsabs.harvard.edu/abs/2019IJMPD..2850154E} {28, 1950154}

\bibitem[\protect\citeauthoryear{{Evans} et~al.,}{{Evans}
  et~al.}{2009}]{Evans2009}
{Evans} P.~A.,  et~al., 2009, \mn@doi [Monthly Notices of the Royal
  Astronomical Society] {10.1111/j.1365-2966.2009.14913.x}, \href
  {https://ui.adsabs.harvard.edu/abs/2009MNRAS.397.1177E} {397, 1177}

\bibitem[\protect\citeauthoryear{{Font-Ribera} et~al.,}{{Font-Ribera}
  et~al.}{2014}]{2014JCAP...05..027F}
{Font-Ribera} A.,  et~al., 2014, \mn@doi [\jcap]
  {10.1088/1475-7516/2014/05/027}, \href
  {https://ui.adsabs.harvard.edu/abs/2014JCAP...05..027F} {2014, 027}

\bibitem[\protect\citeauthoryear{{Freedman}}{{Freedman}}{2021}]{2021ApJ...919...16F}
{Freedman} W.~L.,  2021, \mn@doi [\apj] {10.3847/1538-4357/ac0e95}, \href
  {https://ui.adsabs.harvard.edu/abs/2021ApJ...919...16F} {919, 16}

\bibitem[\protect\citeauthoryear{{G{\'o}mez-Valent} \&
  {Amendola}}{{G{\'o}mez-Valent} \& {Amendola}}{2018}]{2018JCAP...04..051G}
{G{\'o}mez-Valent} A.,  {Amendola} L.,  2018, \mn@doi [\jcap]
  {10.1088/1475-7516/2018/04/051}, \href
  {https://ui.adsabs.harvard.edu/abs/2018JCAP...04..051G} {2018, 051}

\bibitem[\protect\citeauthoryear{{Gonzalez}, {Benetti}, {von Marttens}  \&
  {Alcaniz}}{{Gonzalez} et~al.}{2021}]{2021JCAP...11..060G}
{Gonzalez} J.~E.,  {Benetti} M.,  {von Marttens} R.,   {Alcaniz} J.,  2021,
  \mn@doi [\jcap] {10.1088/1475-7516/2021/11/060}, \href
  {https://ui.adsabs.harvard.edu/abs/2021JCAP...11..060G} {2021, 060}

\bibitem[\protect\citeauthoryear{Hinshaw et~al.,}{Hinshaw
  et~al.}{2013}]{Hinshaw_2013}
Hinshaw G.,  et~al., 2013, \mn@doi [The Astrophysical Journal Supplement
  Series] {10.1088/0067-0049/208/2/19}, 208, 19

\bibitem[\protect\citeauthoryear{{Horowitz} \& {Teukolsky}}{{Horowitz} \&
  {Teukolsky}}{1999}]{1999RvMPS..71..180H}
{Horowitz} G.~T.,  {Teukolsky} S.~A.,  1999, \mn@doi [Reviews of Modern Physics
  Supplement] {10.1103/RevModPhys.71.S180}, \href
  {https://ui.adsabs.harvard.edu/abs/1999RvMPS..71..180H} {71, S180}

\bibitem[\protect\citeauthoryear{{Just}, {Brandt}, {Shemmer}, {Steffen},
  {Schneider}, {Chartas}  \& {Garmire}}{{Just} et~al.}{2007}]{just07}
{Just} D.~W.,  {Brandt} W.~N.,  {Shemmer} O.,  {Steffen} A.~T.,  {Schneider}
  D.~P.,  {Chartas} G.,   {Garmire} G.~P.,  2007, \mn@doi [\apj]
  {10.1086/519990}, \href {http://adsabs.harvard.edu/abs/2007ApJ...665.1004J}
  {665, 1004}

\bibitem[\protect\citeauthoryear{{Kelly}}{{Kelly}}{2007}]{Kelly2007}
{Kelly} B.~C.,  2007, \mn@doi [\apj] {10.1086/519947}, \href
  {https://ui.adsabs.harvard.edu/abs/2007ApJ...665.1489K} {665, 1489}

\bibitem[\protect\citeauthoryear{{Khadka} \& {Ratra}}{{Khadka} \&
  {Ratra}}{2020a}]{2020MNRAS.492.4456K}
{Khadka} N.,  {Ratra} B.,  2020a, \mn@doi [\mnras] {10.1093/mnras/staa101},
  \href {https://ui.adsabs.harvard.edu/abs/2020MNRAS.492.4456K} {492, 4456}

\bibitem[\protect\citeauthoryear{{Khadka} \& {Ratra}}{{Khadka} \&
  {Ratra}}{2020b}]{2020MNRAS.497..263K}
{Khadka} N.,  {Ratra} B.,  2020b, \mn@doi [\mnras] {10.1093/mnras/staa1855},
  \href {https://ui.adsabs.harvard.edu/abs/2020MNRAS.497..263K} {497, 263}

\bibitem[\protect\citeauthoryear{{Khadka} \& {Ratra}}{{Khadka} \&
  {Ratra}}{2021}]{2021MNRAS.502.6140K}
{Khadka} N.,  {Ratra} B.,  2021, \mn@doi [\mnras] {10.1093/mnras/stab486},
  \href {https://ui.adsabs.harvard.edu/abs/2021MNRAS.502.6140K} {502, 6140}

\bibitem[\protect\citeauthoryear{{Khadka} \& {Ratra}}{{Khadka} \&
  {Ratra}}{2022}]{2022MNRAS.510.2753K}
{Khadka} N.,  {Ratra} B.,  2022, \mn@doi [\mnras] {10.1093/mnras/stab3678},
  \href {https://ui.adsabs.harvard.edu/abs/2022MNRAS.510.2753K} {510, 2753}

\bibitem[\protect\citeauthoryear{{Kocevski} \& {Liang}}{{Kocevski} \&
  {Liang}}{2006}]{2006ApJ...642..371K}
{Kocevski} D.,  {Liang} E.,  2006, \mn@doi [\apj] {10.1086/500816}, \href
  {https://ui.adsabs.harvard.edu/abs/2006ApJ...642..371K} {642, 371}

\bibitem[\protect\citeauthoryear{{Kroupa}, {Subr}, {Jerabkova}  \&
  {Wang}}{{Kroupa} et~al.}{2020}]{2020MNRAS.498.5652K}
{Kroupa} P.,  {Subr} L.,  {Jerabkova} T.,   {Wang} L.,  2020, \mn@doi [\mnras]
  {10.1093/mnras/staa2276}, \href
  {https://ui.adsabs.harvard.edu/abs/2020MNRAS.498.5652K} {498, 5652}

\bibitem[\protect\citeauthoryear{{Lenart}, {Bargiacchi}, {Dainotti}, {Nagataki}
   \& {Capozziello}}{{Lenart} et~al.}{2023}]{biasfreeQSO2022}
{Lenart} A.~{\L}.,  {Bargiacchi} G.,  {Dainotti} M.~G.,  {Nagataki} S.,
  {Capozziello} S.,  2023, \mn@doi [\apjs] {10.3847/1538-4365/aca404}, \href
  {https://ui.adsabs.harvard.edu/abs/2023ApJS..264...46L} {264, 46}

\bibitem[\protect\citeauthoryear{{Levine}, {Dainotti}, {Zvonarek}, {Fraija},
  {Warren}, {Chandra}  \& {Lloyd-Ronning}}{{Levine}
  et~al.}{2022}]{Levine2022ApJ...925...15L}
{Levine} D.,  {Dainotti} M.,  {Zvonarek} K.~J.,  {Fraija} N.,  {Warren} D.~C.,
  {Chandra} P.,   {Lloyd-Ronning} N.,  2022, \mn@doi [\apj]
  {10.3847/1538-4357/ac4221}, \href
  {https://ui.adsabs.harvard.edu/abs/2022ApJ...925...15L} {925, 15}

\bibitem[\protect\citeauthoryear{{Li}, {Huang}  \& {Wang}}{{Li}
  et~al.}{2022}]{2022MNRAS.517.1901L}
{Li} Z.,  {Huang} L.,   {Wang} J.,  2022, \mn@doi [\mnras]
  {10.1093/mnras/stac2735}, \href
  {https://ui.adsabs.harvard.edu/abs/2022MNRAS.517.1901L} {517, 1901}

\bibitem[\protect\citeauthoryear{{Liao}, {Shafieloo}, {Keeley}  \&
  {Linder}}{{Liao} et~al.}{2019}]{2019ApJ...886L..23L}
{Liao} K.,  {Shafieloo} A.,  {Keeley} R.~E.,   {Linder} E.~V.,  2019, \mn@doi
  [\apjl] {10.3847/2041-8213/ab5308}, \href
  {https://ui.adsabs.harvard.edu/abs/2019ApJ...886L..23L} {886, L23}

\bibitem[\protect\citeauthoryear{{Lusso} \& {Risaliti}}{{Lusso} \&
  {Risaliti}}{2016}]{lr16}
{Lusso} E.,  {Risaliti} G.,  2016, \mn@doi [\apj]
  {10.3847/0004-637X/819/2/154}, \href
  {http://adsabs.harvard.edu/abs/2016ApJ...819..154L} {819, 154}

\bibitem[\protect\citeauthoryear{{Lusso} et~al.,}{{Lusso}
  et~al.}{2010}]{2010A&A...512A..34L}
{Lusso} E.,  et~al., 2010, \mn@doi [\aap] {10.1051/0004-6361/200913298}, \href
  {https://ui.adsabs.harvard.edu/abs/2010A&A...512A..34L} {512, A34}

\bibitem[\protect\citeauthoryear{{Lusso}, {Piedipalumbo}, {Risaliti},
  {Paolillo}, {Bisogni}, {Nardini}  \& {Amati}}{{Lusso}
  et~al.}{2019}]{lusso2019}
{Lusso} E.,  {Piedipalumbo} E.,  {Risaliti} G.,  {Paolillo} M.,  {Bisogni} S.,
  {Nardini} E.,   {Amati} L.,  2019, \mn@doi [\aap]
  {10.1051/0004-6361/201936223}, \href
  {https://ui.adsabs.harvard.edu/abs/2019A&A...628L...4L} {628, L4}

\bibitem[\protect\citeauthoryear{{Lusso} et~al.,}{{Lusso}
  et~al.}{2020}]{2020A&A...642A.150L}
{Lusso} E.,  et~al., 2020, \mn@doi [\aap] {10.1051/0004-6361/202038899}, \href
  {https://ui.adsabs.harvard.edu/abs/2020A&A...642A.150L} {642, A150}

\bibitem[\protect\citeauthoryear{{Netzer}}{{Netzer}}{2013}]{qsophysics}
{Netzer} H.,  2013, \mn@doi [Cambridge University Press]
  {https://doi.org/10.1017/CBO9781139109291}

\bibitem[\protect\citeauthoryear{{Peebles} \& {Ratra}}{{Peebles} \&
  {Ratra}}{2003}]{2003RvMP...75..559P}
{Peebles} P.~J.,  {Ratra} B.,  2003, \mn@doi [Reviews of Modern Physics]
  {10.1103/RevModPhys.75.559}, \href
  {https://ui.adsabs.harvard.edu/abs/2003RvMP...75..559P} {75, 559}

\bibitem[\protect\citeauthoryear{{Perlmutter} et~al.,}{{Perlmutter}
  et~al.}{1999}]{perlmutter1999}
{Perlmutter} S.,  et~al., 1999, \mn@doi [\apj] {10.1086/307221}, \href
  {https://ui.adsabs.harvard.edu/abs/1999ApJ...517..565P} {517, 565}

\bibitem[\protect\citeauthoryear{{Petrosian}, {Bouvier}  \& {Ryde}}{{Petrosian}
  et~al.}{2009}]{2009arXiv0909.5051P}
{Petrosian} V.,  {Bouvier} A.,   {Ryde} F.,  2009, \mn@doi [arXiv e-prints]
  {10.48550/arXiv.0909.5051}, \href
  {https://ui.adsabs.harvard.edu/abs/2009arXiv0909.5051P} {p. arXiv:0909.5051}

\bibitem[\protect\citeauthoryear{{Planck Collaboration} et~al.,}{{Planck
  Collaboration} et~al.}{2020}]{planck2018}
{Planck Collaboration} et~al., 2020, \mn@doi [\aap]
  {10.1051/0004-6361/201833910}, \href
  {https://ui.adsabs.harvard.edu/abs/2020A&A...641A...6P} {641, A6}

\bibitem[\protect\citeauthoryear{{Postnikov}, {Dainotti}, {Hernandez}  \&
  {Capozziello}}{{Postnikov} et~al.}{2014}]{Postnikov14}
{Postnikov} S.,  {Dainotti} M.~G.,  {Hernandez} X.,   {Capozziello} S.,  2014,
  \mn@doi [\apj] {10.1088/0004-637X/783/2/126}, \href
  {http://adsabs.harvard.edu/abs/2014ApJ...783..126P} {783, 126}

\bibitem[\protect\citeauthoryear{{Rea}, {Gull{\'o}n}, {Pons}, {Perna},
  {Dainotti}, {Miralles}  \& {Torres}}{{Rea}
  et~al.}{2015}]{Rea2015ApJ...813...92R}
{Rea} N.,  {Gull{\'o}n} M.,  {Pons} J.~A.,  {Perna} R.,  {Dainotti} M.~G.,
  {Miralles} J.~A.,   {Torres} D.~F.,  2015, \mn@doi [\apj]
  {10.1088/0004-637X/813/2/92}, \href
  {https://ui.adsabs.harvard.edu/abs/2015ApJ...813...92R} {813, 92}

\bibitem[\protect\citeauthoryear{{Rezaei}, {Pour-Ojaghi}  \&
  {Malekjani}}{{Rezaei} et~al.}{2020}]{2020ApJ...900...70R}
{Rezaei} M.,  {Pour-Ojaghi} S.,   {Malekjani} M.,  2020, \mn@doi [\apj]
  {10.3847/1538-4357/aba517}, \href
  {https://ui.adsabs.harvard.edu/abs/2020ApJ...900...70R} {900, 70}

\bibitem[\protect\citeauthoryear{{Riess} et~al.,}{{Riess}
  et~al.}{1998}]{riess1998}
{Riess} A.~G.,  et~al., 1998, \mn@doi [\aj] {10.1086/300499}, \href
  {https://ui.adsabs.harvard.edu/abs/1998AJ....116.1009R} {116, 1009}

\bibitem[\protect\citeauthoryear{{Riess} et~al.,}{{Riess}
  et~al.}{2004}]{2004ApJ...607..665R}
{Riess} A.~G.,  et~al., 2004, \mn@doi [\apj] {10.1086/383612}, \href
  {https://ui.adsabs.harvard.edu/abs/2004ApJ...607..665R} {607, 665}

\bibitem[\protect\citeauthoryear{{Riess} et~al.,}{{Riess}
  et~al.}{2022}]{2022ApJ...934L...7R}
{Riess} A.~G.,  et~al., 2022, \mn@doi [\apjl] {10.3847/2041-8213/ac5c5b}, \href
  {https://ui.adsabs.harvard.edu/abs/2022ApJ...934L...7R} {934, L7}

\bibitem[\protect\citeauthoryear{{Risaliti} \& {Lusso}}{{Risaliti} \&
  {Lusso}}{2015}]{rl15}
{Risaliti} G.,  {Lusso} E.,  2015, \mn@doi [\apj] {10.1088/0004-637X/815/1/33},
  \href {http://adsabs.harvard.edu/abs/2015ApJ...815...33R} {815, 33}

\bibitem[\protect\citeauthoryear{{Risaliti} \& {Lusso}}{{Risaliti} \&
  {Lusso}}{2019}]{rl19}
{Risaliti} G.,  {Lusso} E.,  2019, \mn@doi [Nature Astronomy]
  {10.1038/s41550-018-0657-z}, \href
  {https://ui.adsabs.harvard.edu/abs/2019NatAs.tmp..195R} {p.~195}

\bibitem[\protect\citeauthoryear{{Rodney} et~al.,}{{Rodney}
  et~al.}{2015}]{Rodney}
{Rodney} S.~A.,  et~al., 2015, \mn@doi [\aj] {10.1088/0004-6256/150/5/156},
  \href {https://ui.adsabs.harvard.edu/abs/2015AJ....150..156R} {150, 156}

\bibitem[\protect\citeauthoryear{{Rowlinson}, {Gompertz}, {Dainotti},
  {O'Brien}, {Wijers}  \& {van der Horst}}{{Rowlinson}
  et~al.}{2014}]{Rowlinson2014MNRAS.443.1779R}
{Rowlinson} A.,  {Gompertz} B.~P.,  {Dainotti} M.,  {O'Brien} P.~T.,  {Wijers}
  R.~A.~M.~J.,   {van der Horst} A.~J.,  2014, \mn@doi [\mnras]
  {10.1093/mnras/stu1277}, \href
  {https://ui.adsabs.harvard.edu/abs/2014MNRAS.443.1779R} {443, 1779}

\bibitem[\protect\citeauthoryear{{Sahni} \& {Starobinsky}}{{Sahni} \&
  {Starobinsky}}{2000}]{2000IJMPD...9..373S}
{Sahni} V.,  {Starobinsky} A.,  2000, \mn@doi [International Journal of Modern
  Physics D] {10.1142/S0218271800000542}, \href
  {https://ui.adsabs.harvard.edu/abs/2000IJMPD...9..373S} {9, 373}

\bibitem[\protect\citeauthoryear{Schiavone, Montani  \& Bombacigno}{Schiavone
  et~al.}{2022}]{Schiavone:2022wvq}
Schiavone T.,  Montani G.,   Bombacigno F.,  2022

\bibitem[\protect\citeauthoryear{{Scolnic} et~al.,}{{Scolnic}
  et~al.}{2018}]{scolnic2018}
{Scolnic} D.~M.,  et~al., 2018, \mn@doi [\apj] {10.3847/1538-4357/aab9bb},
  \href {https://ui.adsabs.harvard.edu/abs/2018ApJ...859..101S} {859, 101}

\bibitem[\protect\citeauthoryear{{Scolnic} et~al.,}{{Scolnic}
  et~al.}{2022}]{pantheon+}
{Scolnic} D.,  et~al., 2022, \mn@doi [\apj] {10.3847/1538-4357/ac8b7a}, \href
  {https://ui.adsabs.harvard.edu/abs/2022ApJ...938..113S} {938, 113}

\bibitem[\protect\citeauthoryear{{Sharov}}{{Sharov}}{2016}]{2016JCAP...06..023S}
{Sharov} G.~S.,  2016, \mn@doi [\jcap] {10.1088/1475-7516/2016/06/023}, \href
  {https://ui.adsabs.harvard.edu/abs/2016JCAP...06..023S} {2016, 023}

\bibitem[\protect\citeauthoryear{{Sharov} \& {Vasiliev}}{{Sharov} \&
  {Vasiliev}}{2018}]{2018arXiv180707323S}
{Sharov} G.~S.,  {Vasiliev} V.~O.,  2018, arXiv e-prints, \href
  {https://ui.adsabs.harvard.edu/abs/2018arXiv180707323S} {p. arXiv:1807.07323}

\bibitem[\protect\citeauthoryear{{Singal}, {Petrosian}, {Lawrence}  \&
  {Stawarz}}{{Singal} et~al.}{2011}]{2011ApJ...743..104S}
{Singal} J.,  {Petrosian} V.,  {Lawrence} A.,   {Stawarz} {\L}.,  2011, \mn@doi
  [\apj] {10.1088/0004-637X/743/2/104}, \href
  {https://ui.adsabs.harvard.edu/abs/2011ApJ...743..104S} {743, 104}

\bibitem[\protect\citeauthoryear{{Srianand} \& {Gopal-Krishna}}{{Srianand} \&
  {Gopal-Krishna}}{1998}]{1998A&A...334...39S}
{Srianand} R.,  {Gopal-Krishna} 1998, \aap, \href
  {https://ui.adsabs.harvard.edu/abs/1998A&A...334...39S} {334, 39}

\bibitem[\protect\citeauthoryear{Srinivasaragavan, Dainotti, Fraija, Hernandez,
  Nagataki, Lenart, Bowden  \& Wagner}{Srinivasaragavan
  et~al.}{2020}]{Srinivasaragavan2020}
Srinivasaragavan G.~P.,  Dainotti M.~G.,  Fraija N.,  Hernandez X.,  Nagataki
  S.,  Lenart A.,  Bowden L.,   Wagner R.,  2020, \mn@doi [The Astrophysical
  Journal] {10.3847/1538-4357/abb702}, 903, 18

\bibitem[\protect\citeauthoryear{{Steffen}, {Strateva}, {Brandt}, {Alexander},
  {Koekemoer}, {Lehmer}, {Schneider}  \& {Vignali}}{{Steffen}
  et~al.}{2006}]{steffen06}
{Steffen} A.~T.,  {Strateva} I.,  {Brandt} W.~N.,  {Alexander} D.~M.,
  {Koekemoer} A.~M.,  {Lehmer} B.~D.,  {Schneider} D.~P.,   {Vignali} C.,
  2006, \mn@doi [\aj] {10.1086/503627}, \href
  {http://adsabs.harvard.edu/abs/2006AJ....131.2826S} {131, 2826}

\bibitem[\protect\citeauthoryear{{Tananbaum} et~al.,}{{Tananbaum}
  et~al.}{1979}]{1979ApJ...234L...9T}
{Tananbaum} H.,  et~al., 1979, \mn@doi [\apjl] {10.1086/183100}, \href
  {https://ui.adsabs.harvard.edu/abs/1979ApJ...234L...9T} {234, L9}

\bibitem[\protect\citeauthoryear{{Tripp}}{{Tripp}}{1998}]{1998A&A...331..815T}
{Tripp} R.,  1998, \aap, \href
  {https://ui.adsabs.harvard.edu/abs/1998A&A...331..815T} {331, 815}

\bibitem[\protect\citeauthoryear{{Vagnozzi}, {Loeb}  \& {Moresco}}{{Vagnozzi}
  et~al.}{2021}]{2021ApJ...908...84V}
{Vagnozzi} S.,  {Loeb} A.,   {Moresco} M.,  2021, \mn@doi [\apj]
  {10.3847/1538-4357/abd4df}, \href
  {https://ui.adsabs.harvard.edu/abs/2021ApJ...908...84V} {908, 84}

\bibitem[\protect\citeauthoryear{{Visser}}{{Visser}}{2004}]{2004CQGra..21.2603V}
{Visser} M.,  2004, \mn@doi [Classical and Quantum Gravity]
  {10.1088/0264-9381/21/11/006}, \href
  {https://ui.adsabs.harvard.edu/abs/2004CQGra..21.2603V} {21, 2603}

\bibitem[\protect\citeauthoryear{{Wang} et~al.,}{{Wang}
  et~al.}{2021}]{2021ApJ...907L...1W}
{Wang} F.,  et~al., 2021, \mn@doi [\apjl] {10.3847/2041-8213/abd8c6}, \href
  {https://ui.adsabs.harvard.edu/abs/2021ApJ...907L...1W} {907, L1}

\bibitem[\protect\citeauthoryear{{Wang}, {Liu}, {Yuan}, {Liang}, {Yu}  \&
  {Wu}}{{Wang} et~al.}{2022}]{2022arXiv221014432W}
{Wang} B.,  {Liu} Y.,  {Yuan} Z.,  {Liang} N.,  {Yu} H.,   {Wu} P.,  2022,
  arXiv e-prints, \href {https://ui.adsabs.harvard.edu/abs/2022arXiv221014432W}
  {p. arXiv:2210.14432}

\bibitem[\protect\citeauthoryear{{Weinberg}}{{Weinberg}}{1972}]{1972gcpa.book.....W}
{Weinberg} S.,  1972, {Gravitation and Cosmology: Principles and Applications
  of the General Theory of Relativity}

\bibitem[\protect\citeauthoryear{{Weinberg}}{{Weinberg}}{1989}]{1989RvMP...61....1W}
{Weinberg} S.,  1989, \mn@doi [Reviews of Modern Physics]
  {10.1103/RevModPhys.61.1}, \href
  {https://ui.adsabs.harvard.edu/abs/1989RvMP...61....1W} {61, 1}

\bibitem[\protect\citeauthoryear{{Willingale} et~al.,}{{Willingale}
  et~al.}{2007}]{2007ApJ...662.1093W}
{Willingale} R.,  et~al., 2007, \mn@doi [\apj] {10.1086/517989}, \href
  {https://ui.adsabs.harvard.edu/abs/2007ApJ...662.1093W} {662, 1093}

\bibitem[\protect\citeauthoryear{{Wong} et~al.,}{{Wong}
  et~al.}{2020}]{2020MNRAS.498.1420W}
{Wong} K.~C.,  et~al., 2020, \mn@doi [\mnras] {10.1093/mnras/stz3094}, \href
  {https://ui.adsabs.harvard.edu/abs/2020MNRAS.498.1420W} {498, 1420}

\bibitem[\protect\citeauthoryear{{Zamora Mun{\~o}z} \&
  {Escamilla-Rivera}}{{Zamora Mun{\~o}z} \&
  {Escamilla-Rivera}}{2020}]{2020JCAP...12..007Z}
{Zamora Mun{\~o}z} C.,  {Escamilla-Rivera} C.,  2020, \mn@doi [\jcap]
  {10.1088/1475-7516/2020/12/007}, \href
  {https://ui.adsabs.harvard.edu/abs/2020JCAP...12..007Z} {2020, 007}

\bibitem[\protect\citeauthoryear{{Zamorani} et~al.,}{{Zamorani}
  et~al.}{1981}]{1981ApJ...245..357Z}
{Zamorani} G.,  et~al., 1981, \mn@doi [\apj] {10.1086/158815}, \href
  {https://ui.adsabs.harvard.edu/abs/1981ApJ...245..357Z} {245, 357}

\bibitem[\protect\citeauthoryear{{Zhang} \& {M{\'e}sz{\'a}ros}}{{Zhang} \&
  {M{\'e}sz{\'a}ros}}{2001}]{2001ApJ...552L..35Z}
{Zhang} B.,  {M{\'e}sz{\'a}ros} P.,  2001, \mn@doi [\apjl] {10.1086/320255},
  \href {https://ui.adsabs.harvard.edu/abs/2001ApJ...552L..35Z} {552, L35}

\makeatother
\end{thebibliography}







\bsp	
\label{lastpage}
\end{document}